\documentclass[twocolumn]{aastex631}
\usepackage[utf8]{inputenc}
\usepackage{amsmath, amsthm, amssymb}
\usepackage{epstopdf}
\usepackage{graphicx}
\usepackage{float}
\usepackage{subfigure}
\usepackage{indentfirst}
\usepackage{color}
\usepackage{natbib}
\usepackage[figuresright]{rotating}
\usepackage{enumerate}

\makeatletter

\newcommand{\Rmnum}[1]{\expandafter\@slowromancap\romannumeral #1@}
\makeatother

\begin{document}

    \title{Galactic Cirri at High Galactic Latitudes: \Rmnum{1}. Investigating Scatter in Slopes between Optical and far-Infrared Intensities}

    \author[0009-0000-6954-9825]{Yunning Zhao}
    \affiliation{CAS Key Laboratory of Optical Astronomy, National Astronomical Observatories,  Chinese Academy of Sciences, Beijing 100101, People's Republic of China}
    \affiliation{School of Astronomy and Space Science, University of Chinese Academy of Sciences, Beijing 100049, China}

    \author[0000-0002-1783-957X]{Wei Zhang{$^\dagger$}}
    \affiliation{CAS Key Laboratory of Optical Astronomy, National Astronomical Observatories,  Chinese Academy of Sciences, Beijing 100101, People's Republic of China}
    \email{Wei Zhang{$^\dagger$}, xtwfn@bao.ac.cn}
    
    \author{Lin Ma}
    \affiliation{{The Key Laboratory of Cosmic Rays (Tibet University), Ministry of Education, Lhasa 850000, Tibet, China}}
    \affiliation{CAS Key Laboratory of Optical Astronomy, National Astronomical Observatories,  Chinese Academy of Sciences, Beijing 100101, People's Republic of China}
    \affiliation{School of Astronomy and Space Science, University of Chinese Academy of Sciences, Beijing 100049, China}
    
    \author[0009-0008-1361-4825]{Shiming Wen}
    \affiliation{CAS Key Laboratory of Optical Astronomy, National Astronomical Observatories,  Chinese Academy of Sciences, Beijing 100101, People's Republic of China}
    
    \author{Hong Wu{$^\ddagger$}}
    \affiliation{CAS Key Laboratory of Optical Astronomy, National Astronomical Observatories,  Chinese Academy of Sciences, Beijing 100101, People's Republic of China}
    \email{Hong Wu{$^\ddagger$}, hwu@bao.ac.cn}

\begin{abstract}

    Based on the slopes between DESI $g,r$ and IRAS 100 $\mu m$ intensities, specifically $k_{g}$ and $k_{r}$, we have constructed a substantial sample of Galactic cirri. This sample covers 561.25 deg$^2$ at high Galactic latitudes ($|b| \geq 30\degr$),  allowing for a systematic study of the physical parameters of the Galactic cirrus on a large scale, such as $g-r$ color, dust temperature, asymmetry factor and albedo. The ratio of $k_{g}$ and $k_{r}$ is consistent with the diffuse Galactic starlight model, suggesting that the diffuse starlight within our own Galaxy serves as the primary illumination source for the cirrus. Both $k_{g}$ and $k_{r}$ decrease slowly with increasing Galactic latitudes and IRAS 100 $\mu m$ intensities, while they do not have a correlation with Galactic longitudes. The distribution of $k_{g}$ and $k_{r}$ confirms a significant scatter in the slopes, reaching a factor of 4--5. Such large scatter cannot be explained by the weak correlation between the slopes and Galactic latitudes and IRAS 100 $\mu m$ intensities. Instead, it is attributed to substantial variations in the intrinsic properties of the dust, e.g., asymmetry factor and albedo. We propose that the properties of dust particles play a critical role in the observed scatter in slopes, making them the primary contributing factors. Moreover, the variations in dust properties within the cirrus are localized rather than exhibiting large-scale gradients.
    
\end{abstract}

\section{Introduction}

    Galactic cirrus was first quantitatively observed near star-forming regions by \citet{elvey_photoelectric_1937}. They identified an excess of light in addition to the zodiacal light and airglow when measuring the night sky, and they hypothesized that this extra starlight might originate from scattered light by dust. \citet{sandage_high-latitude_1976} discovered Galactic cirri at high Galactic latitudes, and calculations indicated that these reflection nebulae were illuminated by the integrated light from the Galactic plane. The infrared maps obtained by the Infrared Astronomical Satellite (IRAS) were the first to reveal extended far-infrared emissions at high Galactic latitudes, uncovering the interstellar dust emission known as ``infrared cirrus" \citep{low_infrared_1984}.

    Since then, Galactic cirrus has been observed across a wide range of wavelengths, from ultraviolet (UV) to infrared (IR). In the UV, \citet{witt_radiative_1997} discovered that the dominant component contributing to the measured far-UV background is the Diffuse Galactic Light (DGL), which results from scattered light by interstellar dust. \citet{boissier_galex_2015} detected cirrus within the Virgo cluster using GALEX data and obtained diffuse FUV emission maps which could be used to study dust properties. In the optical range, the current generation of deep optical surveys has made it easier to detect Galactic cirrus. \citet{roman_galactic_2020} detected cirrus in the Sloan Digital Sky Survey (SDSS) Stripe82 region \citep{abazajian_seventh_2009} and found that the optical colors of Galactic cirrus are bluer than those of extragalactic sources. Additionally, \citet{zhang_mysterious_2020} discovered a ring-like cirrus in the Beijing–Arizona Sky Survey (BASS) and discussed its properties using multi-band images. \citet{smirnov_prospects_2023} adopted a new method to identify cirrus using machine learning and neural networks, which proves the possibility of machine learning methods to solve this problem. In the far-IR, Galactic cirrus exhibits its peak emission due to its low temperature, which is approximately 20 K \citep{low_infrared_1984, veneziani_properties_2010}. Therefore, far-IR images are commonly employed to identify the presence of cirrus, such as full-sky far-IR maps produced by IRIS \citep{miville-deschenes_iris_2005}. The ESA Herschel Space Observatory also provided far-IR images with higher spatial resolution, allowing for more accurate identification of cirri, although the sky coverage of $Herschel$ map is limited \citep{mihos_burrell_2016}. In the near-IR, the detection of cirrus is relatively more challenging due to its faintness \citep{arendt_cobe_1998, matsumoto_reanalysis_2015}. \citet{sano_derivation_2015} detected the DGL component at high Galactic latitudes using data from the Diffuse Infrared Background Experiment (DIRBE) at 1.25 and 2.2 $\rm \mu m$. Following the detection of dust emission in IR, atomic gas associated with dust grains has also been observed within cirrus \citep{boulanger_diffuse_1988}. These HI clouds exhibit two distinct gas components: low-velocity clouds (LVC) and intermediate-velocity clouds (IVC). Notably, the properties of dust within the IVC differ from those observed in the LVC \citep{bianchi_herschel_2017}.
    
    Another approach to understanding cirrus involves utilizing blank sky spectra to derive the spectrum of DGL. \citet{brandt_spectrum_2012} employed blank sky calibration spectra taken by SDSS-II to measure the optical spectrum of DGL, and the spectral features were found to be consistent with scattered starlight. Subsequently, \citet{kawara_ultraviolet_2017} further decomposed the DGL spectrum by analyzing HST/FOS blank sky spectra spanning the UV to optical wavelengths (0.2-0.7 $\mu m$). More recently, \citet{chellew_optical_2022} obtained a similar optical spectrum of DGL using blank sky spectra from SDSS-III/BOSS. They observed a difference in the DGL continuum between regions in the northern and southern Galactic hemispheres.
 
    The correlation between cirrus intensities in different bands has been an area of interest, particularly the correlation between optical and far-IR bands. \citet{de_vries_comparison_1985} initially compared optical and far-IR images of cirrus and discovered a strong correlation between them. When measuring the DGL optical spectrum, \citet{brandt_spectrum_2012} utilized the robust correlation between optical and far-IR data, which proved to be a valuable tool for distinguishing cirrus. Under optically thin conditions, \citet{ienaka_diffuse_2013} analyzed the cloud MBM32 at high Galactic latitude and observed a linear correlation between DGL and far-IR brightness. Similarly, a linear relationship was observed in the ring-like nebula discovered by \citet{zhang_mysterious_2020}.
    
    In addition, the correlations among cirrus intensities in the UV, near-IR, far-IR, and HI bands have been demonstrated. 
    \citet{boissier_galex_2015} examined the correlation between FUV, NUV, and IR. They identified the potential of FUV images as dust tracers due to their good resolution and sensitivity. The conclusion drawn was that FUV images could serve as an important complementary to traditional dust tracers.
    \citet{akshaya_diffuse_2018} examined the correlation between FUV, NUV, and far-IR emissions at the north and south Galactic poles, focusing their study on diffuse dust-scattered light and extragalactic radiation. Regarding the correlation between near-IR and far-IR, \citet{sano_first_2016}  computed intensity ratios of near-IR DGL to far-IR dust emission and found that these ratios steeply decrease toward high Galactic latitudes. Furthermore, \citet{onishi_miris_2018} demonstrated linear correlations between near-IR DGL and 100 $\mu m$ emission, which is consistent with the findings reported by \citet{sano_derivation_2015}. Additionally, they were the first to derive the linear correlation between near-IR and optical DGL. There is also a tight correlation observed between FIR intensity and HI column density. The emissivity, which represents the cirrus surface brightness per HI column density, is used to constraint the dust model properties \citep{planck_collaboration_planck_2014, bianchi_herschel_2017}. Moreover, a correlation has been observed between molecular CO and H$_{2}$ lines. Overall, researchers have investigated and established correlations among cirrus intensities in various bands, including the UV, near-IR, far-IR, and HI. These findings provide valuable insights into the relationships and properties of different components, such as neutral hydrogen and dust within cirrus regions.
 
    There is currently an issue concerning the correlation between multi-band intensities, specifically, the fitting slopes exhibit significant scatter in different fields, with differences of up to a factor of 3-4. Previous studies have attempted to address this issue from various perspectives. \citet{ienaka_diffuse_2013} compiled fitting slopes from their own sample as well as from previous studies. They discovered that variations in optical depth can modify the fitting slopes by a factor of 2-3. In addition, the interplay between properties of dust grains (such as asymmetry factor and albedo) and the non-isotropic interstellar radiation field (ISRF) also contributes to the scatter in slopes. Furthermore, \citet{sano_first_2016} were the first to detect the dependence of near-IR DGL on Galactic latitude. However, the current research samples exploring this issue are relatively small, usually covering only a dozen square degrees.

    Our aim of this study and subsequent publications is to analyze the multi-band characteristics of large cirrus samples, which encompass UV, optical, IR, and HI wavelengths. In the first of a series of works, we propose to utilize a combination of DESI optical images and IRAS far-IR data to identify Galactic cirrus, generate a substantial sample, and investigate the factors responsible for the scatter in fitting slopes observed between optical and far-IR bands. Thanks to the extensive sky coverage and deep detection capabilities offered by DESI imaging surveys, we can effectively identify Galactic cirrus across large regions of the sky in the optical band. In a follow-up paper, we will explore extended red emission (ERE) and analyze the dynamical properties of Galactic cirri using a broader range of data from multi-band.
 
    The paper is structured as follows. We begin by discussing the selected fields and providing a description of the data used in Section \ref{sec:Data}. The process of data reduction and the creation of the large sample are presented in Section \ref{sec:Data reduction and sample construction}.  In Section \ref{sec:Data analysis and results}, we perform calculations of various parameters associated with cirrus, including color, optical effective temperature, dust temperature, asymmetry factor, and albedo. The factors contributing to the scatter in slopes are discussed in Section \ref{sec:Discussion}. Finally, we draw our conclusions in Section \ref{sec:Conclusion}.

\section{Data} \label{sec:Data}

    An effective approach for identifying cirrus involves examining the strong correlation between optical and far-IR emissions, as highlighted in the study by \citet{brandt_spectrum_2012}. In this section, we provide a concise summary of the fields chosen at high Galactic latitudes and outline the data employed in both the optical and far-IR bands.

        \begin{figure*}[htbp]
        \centering
        \includegraphics[width=0.85\linewidth]{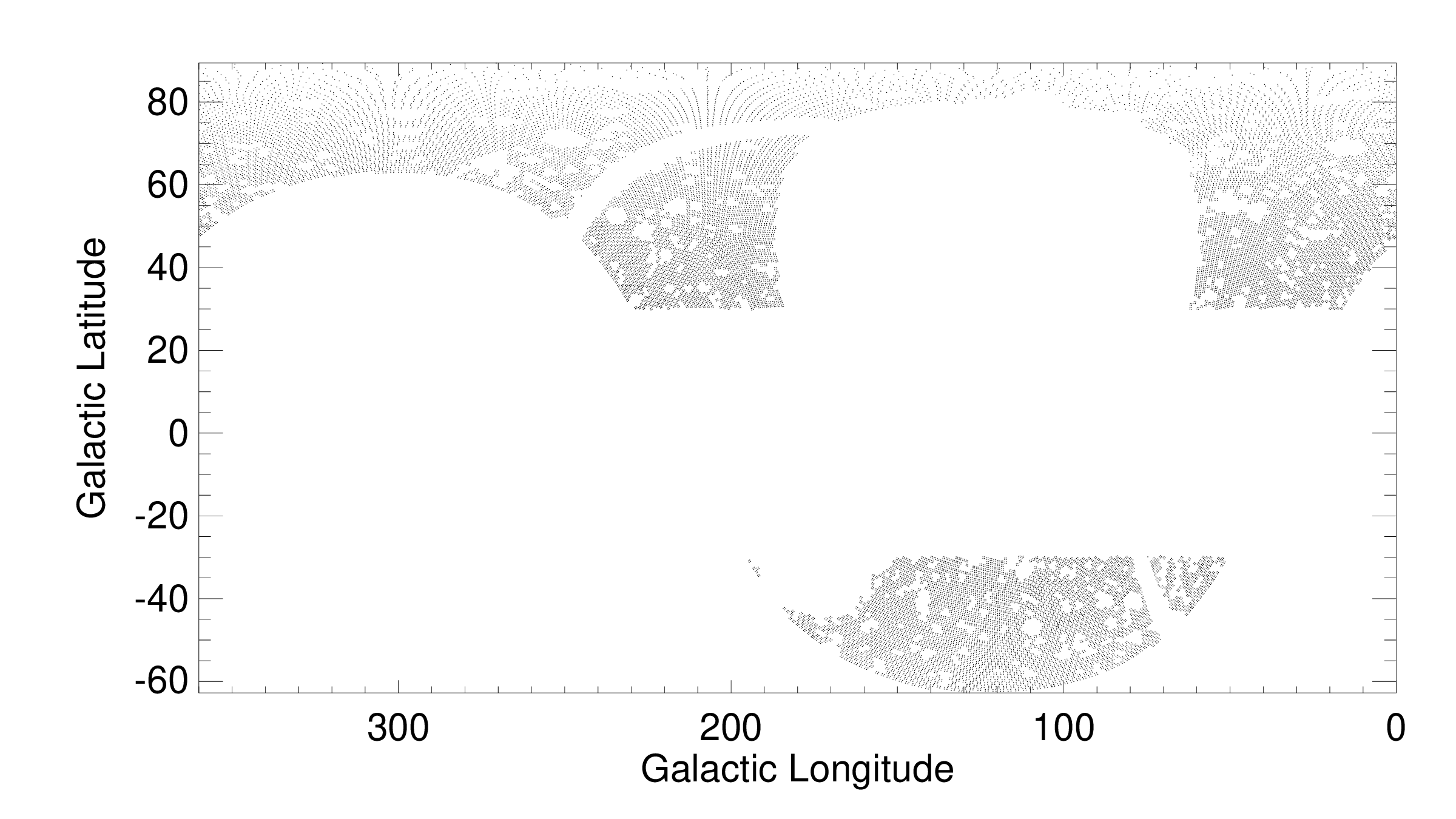}    
        \caption{The sky coverage of the fields that selected from the common regions of DESI, IRAS, GALEX and GALFA-HI surveys at high Galactic latitudes ($\lvert b \rvert \geq 30\degr$). }
        \label{fig:allcoverage}        
        \end{figure*}

    \subsection{Our selected fields}
    
    Considering the study of multi-band properties of cirrus, we have selected regions located at high Galactic latitudes ($\left|b\right| \geq 30\degr$) from the common coverage of various surveys, including the DESI Imaging Legacy Surveys \citep{dey_overview_2019}, the Improved Reprocessing of the IRAS Survey (IRIS) \citep{miville-deschenes_iris_2005}, the Galaxy Evolution Explorer (GALEX) \citep{martin_galaxy_2005}, and the Galactic Arecibo L-band Feed Array HI (GALFA-HI) \citep{peek_galfa-hi_2011}. The reason for selecting regions at high Galactic latitudes is that they are less affected by stars, making it easier to identify fainter Galactic cirrus. Figure \ref{fig:allcoverage} illustrates the coverage of our selected fields across Galactic longitudes ($l$) and Galactic latitudes ($b$). The total coverage area spans 6187 deg$^{2}$, with each point in Figure \ref{fig:allcoverage} representing one field that covers an area of $1\degr \times 1\degr$.

    \subsection{Optical data}

        \begin{figure*}[htbp]
        \centering
        \includegraphics[width=\linewidth]{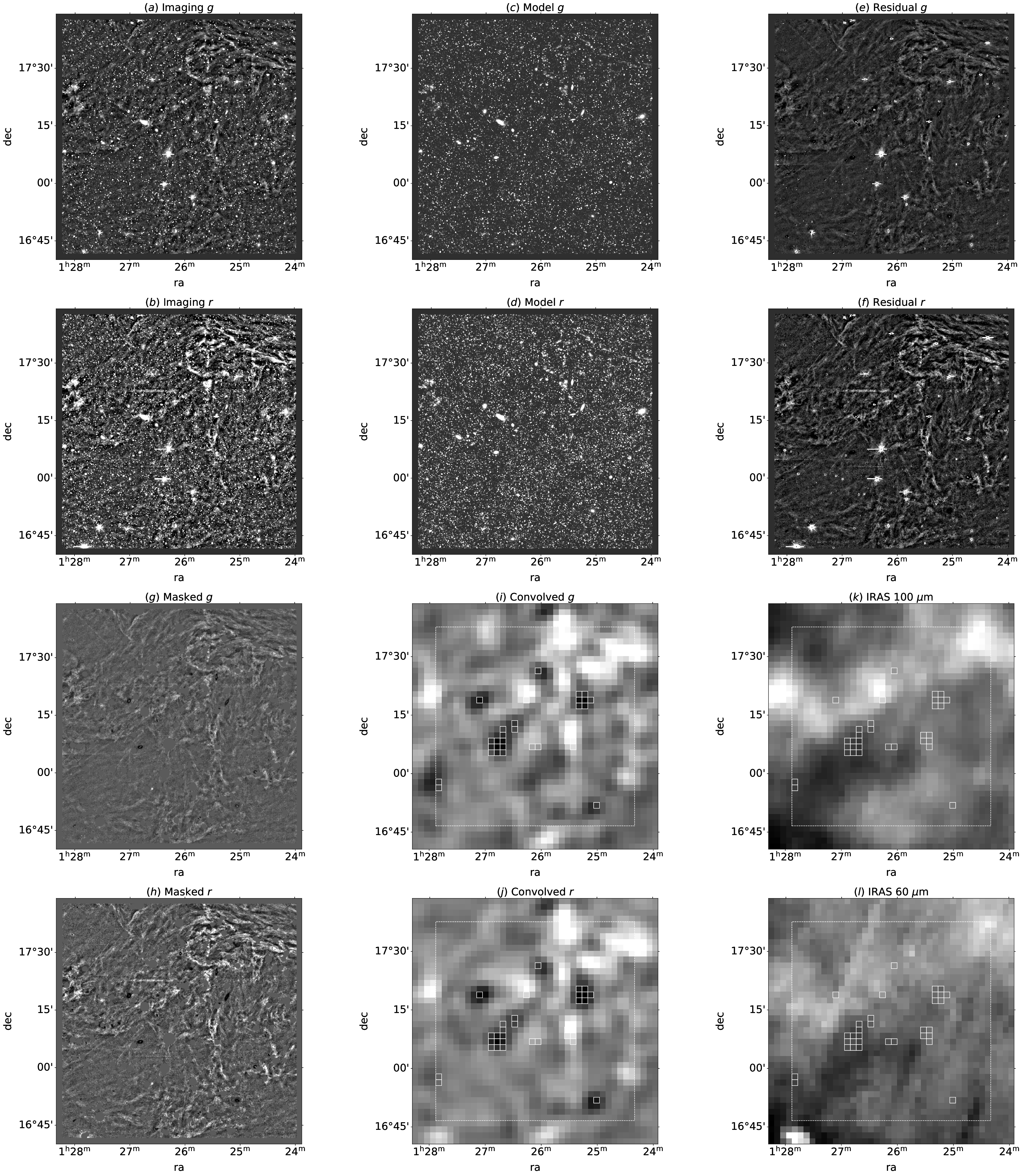}    
        \caption{Example field images covering 1 $\times$ 1 deg$^2$. Panels ($a$) $\sim$ ($f$) display the actual imaging data, model images and residual images in $g$ and $r$ band of DESI Imaging Legacy Surveys. Panels ($g$) and ($h$) are the processed images after being masked in $g$ and $r$ band. Panels ($i$) and ($j$) are the processed images after being convolved and rebinned in $g$ and $r$ band. Panels ($k$) and ($l$) present the IRAS images at 60 and 100 $\mu$m, respectively. In panels ($i$) $\sim$ ($l$), the large dashed white box shows the center three-quarters of the field, and the pixels marked as small solid white boxes are the pixels removed during the iterative process. Pixels within the dashed box that are not marked are reserved for subsequent analysis.}
        \label{fig:0187_resid_process} 
        \end{figure*}

    The original DESI Legacy Imaging Surveys consist of three imaging projects: the Dark Energy Camera Legacy Survey (DECaLS), the Beijing-Arizona Sky Survey (BASS) \citep{zou_third_2019}, and the Mayall z-band Legacy Survey (MzLS). These surveys aim to map the extragalactic sky over an area of 14,000 deg$^{2}$ in three optical bands: $g$, $r$, and $z$. The median 5$\sigma$  point source depths for these bands are 24.0, 23.4, and 22.5 AB mag respectively \citep{dey_overview_2019}. Additionally, the surface brightness limit at a 3$\sigma$ level for a 10 arcsec $\times$ 10 arcsec region can be calculated as follows: $\mu$[3$\sigma$,10$\times$10 arcsec$^2$] = 29.15, 28.73 and 27.52 mag arcsec$^{-2}$ for the $g$, $r$, and $z$ bands respectively \citep{martinez-delgado_giant_2023}. The extensive coverage and high sensitivity of the Legacy Surveys enable the identification of low-surface-brightness structures such as cirrus in large fields. 

    In this work, we use the residual images in the $g$ and $r$ bands of the Data Release 9 (DR9), while excluding the $z$ band due to its relatively poor data quality. It is worth noting that the latest release, Data Release 10 (DR10), incorporates additional DECam data from NOIRLab, which includes the $i$ band in the surveys.
    
    All three types of images, including the actual imaging data, model image, and residual image, can be downloaded from the Legacy Surveys Imagine sky viewer \footnote{\href{https://www.legacysurvey.org/viewer}{https://www.legacysurvey.org/viewer}}. The residual images used in this study are generated by subtracting the model image from the actual imaging data. These residual images have undergone background subtraction, which includes the removal of sky background and zodiacal light. Consequently, most of the bright sources have been effectively subtracted in these residual images, making it easier for us to detect Galactic cirrus. The panels ($a$) $\sim$ ($f$) in Figure \ref{fig:0187_resid_process} display the actual imaging data, model image, and residual image of a specific field in the $g$ and $r$ bands.

    \subsection{Infrared data}
    
    The IRAS provides nearly all-sky images in four infrared bands (12, 25, 60, and 100 $\mu$m). These images were reprocessed by \citet{miville-deschenes_iris_2005} with improvements in zodiacal light subtraction, calibration, compatibility of zero levels, and destriping. This reprocessing effort led to the creation of a new generation of IRAS images, known as the IRIS products. Note that the IRIS images maintain the same pixel scale of $90\arcsec$ as the original IRAS images. 
 
    Since the initial detection of ``infrared cirrus" by \citet{low_infrared_1984}, IRAS 100 $\mu$m images have frequently been used for cirrus identification due to their wide coverage \citep{witt_extended_2008, boissier_galex_2015, roman_galactic_2020}. In this study, we utilize the IRIS products available in the 60 and 100 $\mu$m bands, which can be found on the IRIS Cutouts website\footnote{\href{https://irsa.ipac.caltech.edu/data/IRIS/index\_cutouts.html}{https://irsa.ipac.caltech.edu/data/IRIS/index\_cutouts.html}}. Panels ($k$) and ($l$) of Figure \ref{fig:0187_resid_process} display these images at 60 $\mu$m and 100 $\mu$m, respectively.

\section{Data reduction and sample construction} \label{sec:Data reduction and sample construction}

    \subsection{Image processing} \label{sec:Image processing}
    
    The image processing procedure involves two main procedures: masking bright sources and reducing optical resolution. 

        \subsubsection{Masking bright sources} \label{sec:Masking bright sources}
    
        Although the residual images we used have effectively removed most of the bright sources, some extended galaxies and the halos of bright and saturated stars still persist. These remaining sources can complicate the separation of Galactic cirrus. Therefore, it is necessary to apply masking techniques to eliminate these sources. 
        
        \textbf{Firstly,} for the $g$ and $r$ bands, we employ Gaussian fitting to the flux histogram and remove outliers beyond 3$\sigma$ by setting their values to zero. By this way, we can remove the asterisms from saturated sources, as well as the extreme negative values caused when subtracting backgrounds and models.
        
        \textbf{Next}, we mask bright sources based on a catalog. We download the catalog from Astro data lab \footnote{\href{https://datalab.noirlab.edu/query.php}{https://datalab.noirlab.edu/query.phpl}}. The sources in this catalog are selected to be brighter than 15 mag in the r-band, totaling 2,184,381, which include both stars and galaxies. This catalog contains LS\_ID, type, positions, magnitudes and fluxes in g and r bands for the listed sources, and the half-light radii of galaxy models for galaxy types.
        
\begin{figure}[htbp]
            \centering
            \includegraphics[width=\linewidth]{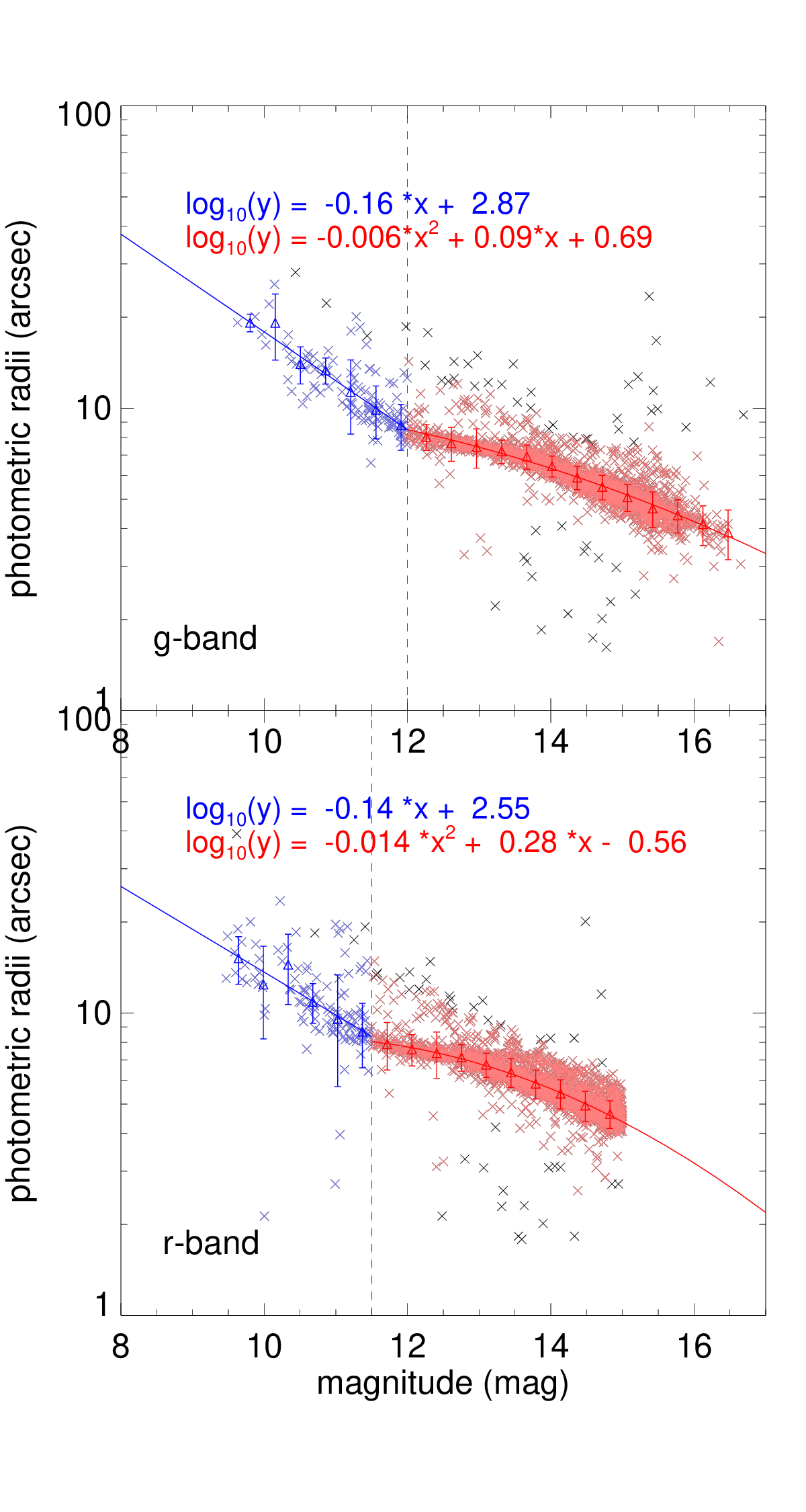}
            \caption{Relationship between magnitude and photometric radius. In each panel, the stars are divided into bins based on x-axis. Triangles represent median values of each bin, accompanied by measurement error bars. Black crosses are points that deviated from the median by more than 3$\sigma$ and are not involved in the subsequent fitting. Blue and red crosses indicate the two groups of bright stars. The vertical dashed lines represent the boundary between the two groups, set at 12 and 11.5 mag for the g and r bands, respectively. Blue and red lines represent the fitting outcomes for median values, with the fitting results displayed at the top of the panels.}
            \label{fig:mag_radii}        
            \end{figure}

        For galaxies in the catalog, we employ a mask radius equal to 5 times the half-light radius. Regarding stars in the catalog, we randomly select 9 fields, which include 2,262 stars, to establish a relationship between magnitude and photometric radius. Photometry is conducted on on their corresponding imaging data using \texttt{SExtractor} \citep{bertin_sextractor_1996}. For each star, we obtain the \texttt{ISOAREA\_IMAGE}, representing the area it occupies in the imaging data measured in units of pixel$^2$. The photometric radius of each star can be calculated by approximating this area as circular.
        We plot the relationship between magnitude and photometric radius for these 2,262 stars in Figure \ref{fig:mag_radii}. These stars are segregated into two groups, with the g and r bands divided by 12 and 11.5 mag respectively. We conduct linear fitting and quadratic fitting for the two groups, represented by the blue and red lines in Figure \ref{fig:mag_radii}. 
        
        We then apply the fitting results to the entire catalog to obtain the photometric radius for all stars. We employ a mask radius equivalent to five times the photometric radius to ensure maximal removal of the effects of bright sources. For g and r bands, we select the larger of the two radii to ensure same masking areas.  The masked images are shown in panels ($g$) and ($h$) of Figure \ref{fig:0187_resid_process}.
        
        \subsubsection{Reducing optical resolution} \label{sec:Reducing optical resolution}
    
        To facilitate correlation analysis between the DESI and IRAS images, it is essential to harmonize the optical resolution with the infrared resolution. The Full Width at Half Maximum (FWHM) of the DESI images is  $1\farcs3$, whereas the FWHM of the IRAS images is $4\farcm3$. To achieve consistency, we convolve the optical images with a Gaussian kernel having a corresponding FWHM ( $\sqrt{ (4.3 \times 60 )^2 - 1.3^2}$ arcseconds) and rebin the pixel size of the optical images (7680 $\times$ 7680) to match the resolution and pixel size of the IR images (42 $\times$ 42). The convolved and rebinned images are displayed in panels ($i$) and ($j$) of Figure \ref{fig:0187_resid_process}, featuring a resolution of $4\farcm3$ and pixel scale of $90\arcsec$.
        We keep the central three-quarters of the field in the subsequent analysis to mitigate the impact of edge blurring caused by Gaussian convolution, as shown by the large dashed white boxes in panels ($i$) and ($j$) of Figure \ref{fig:0187_resid_process}. 
        We can see that there are some remnants mixed with cirrus, and in the next section, we will remove this portion.

    \subsection{Photometry} \label{sec:Photometry}
    
    Following the image processing steps described in Section \ref{sec:Image processing}, we have obtained processed optical images ($g$ and $r$ bands) for a total of 6187 fields, which are accompanied by corresponding IR images at 60 and 100 $\mu$m wavelengths. 
    We divide each field into $2 \times 2$ blocks, with each block covering an area of $0\fdg5 \times 0\fdg5$. In order to remove the effects of bright remnants as mentioned in the \ref{sec:Image processing}, we divide the pixel points of each block into 5 bins based on their $I_{100}$ values and iteratively remove pixel points that deviate from the median by more than 3$\sigma$. The pixel removal process is repeated twice.
    The removed pixels are marked by small solid white boxes in panels ($i$) and ($j$) of Figure \ref{fig:0187_resid_process}. The remaining pixels are used for subsequent analysis.

\begin{figure*}[htbp]
        \centering
        \includegraphics[width=\linewidth,height=\linewidth]{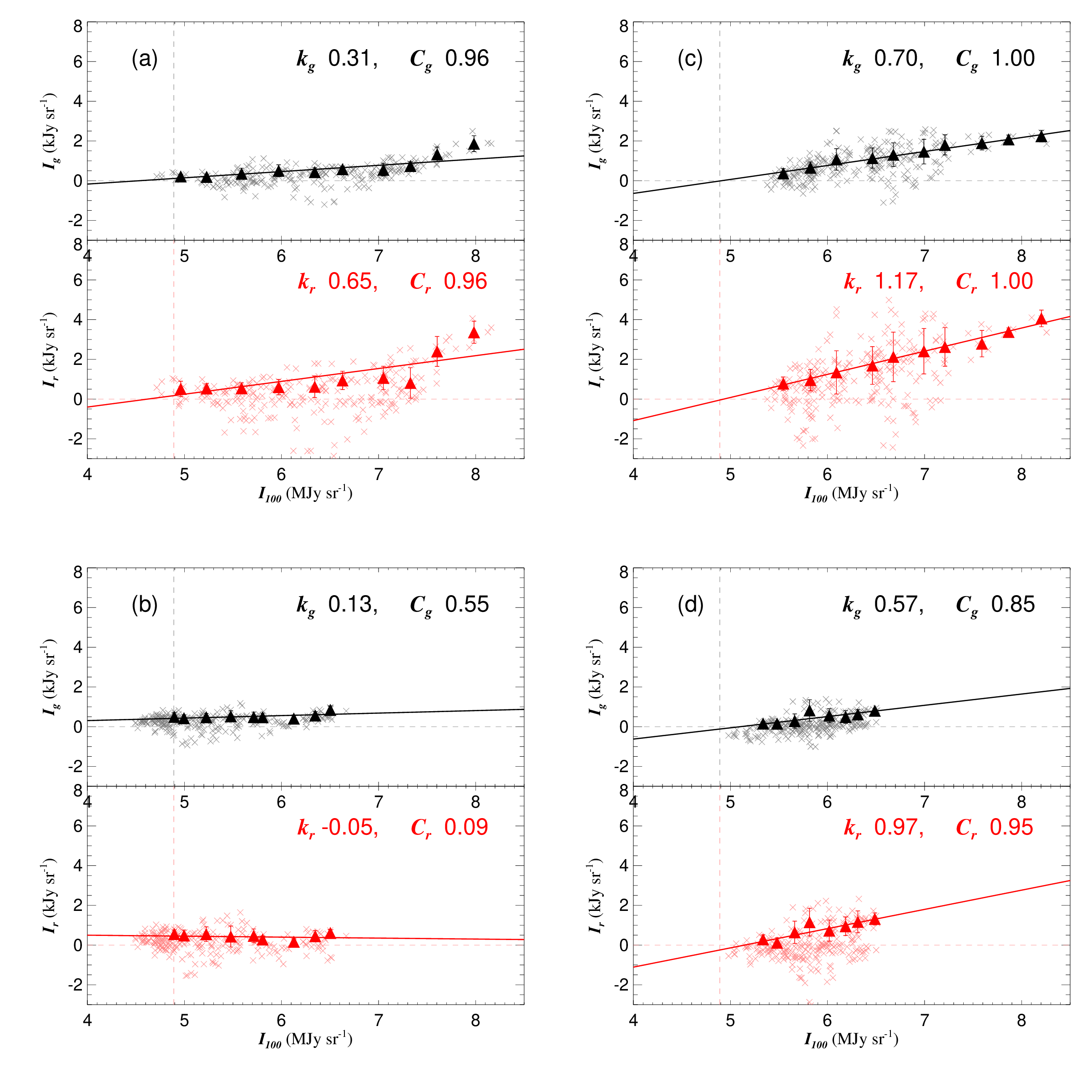}    
        \caption{Correlation between IRAS and DESI intensities is displayed in four blocks, denoted as ($a$),($b$),($c$) and ($d$), representing an example field divided into four blocks. In each panel, the crosses represent the pixels after mask and iterative remove within one block. 
        The horizontal dashed lines represent $I_{g} > 0 $ or $I_{r} > 0 $, while the vertical dashed lines represent $I_{100} > I_{100, limit}$. Only the pixels situated to the right of the vertical lines and above the horizontal lines are considered as cirrus pixels. These cirrus pixels are divided into bins according to $I_{100}$, and the triangles represent the median values in each bin. The solid lines indicate the linear fitting results derived from the median values. Error bars are included to represent the uncertainty in the measurements. In the upper right corner of each panel, the linear fitting slope $k_{\lambda}$ and the correlation coefficients $C_{\lambda}$ are provided.}
        \label{fig:slope_linear}        
        \end{figure*}

    Afterwards, we apply two constraints for the remaining pixels to distinguish cirrus pixels. These constraints are defined as follows: 1) $I_{g} > 0 $ and $I_{r} > 0 $ and 2) $I_{100} > I_{100, limit}$. In Figure \ref{fig:slope_linear}, these constraints are represented by the vertical and horizontal dashed lines. For the first constraint, we discard pixels with optical intensities less than 0, which are considered to be the result of subtracting background excesses and masking remaining sources. For the second constraint, we perform a double Lorentzian fit to the histogram of $I_{100}$  for each field, and set the boundary between the two Lorentzian distributions as $I_{100, limit}$. We regard pixels with intensities smaller than $I_{100, limit}$ as background pixels, while those exceeding $I_{100, limit}$ as cirrus pixels.

    For pixels that satisfy these constrains, we calculate their median intensities ($I_{g,med}$ and $I_{r,med}$), optical surface brightness ($S_{g,med}$ and $S_{r,med}$), and the median 100$\mu$m intensity ($I_{100,med}$). These values serve as the feature values characterizing the cirrus in a single block. Subsequently, we divide the pixels into bins based on $I_{100}$ and calculate the median value. We perform a linear fit and correlation analysis on these median values. As shown in Figure \ref{fig:slope_linear}, the pixels associated with the cirrus exhibit a significant linear correlation between the optical and IR bands. To quantify this correlation, we determine the linear fitting slopes ($k_g, k_r$) and correlation coefficients ($C_g, C_r$) between the intensities of two bands. These values are displayed in the upper right corner of each panel, providing insight into the strength of the relationship between the optical and IR intensities.
    
    \subsection{Sample construction}
  
    Based on the properties observed in Galactic cirrus, we propose the following three selection criteria to select Galactic cirrus sample:
        \begin{enumerate}
            \item $C_{g} \geq 0.5$ and $C_{r} \geq 0.5$
            \item $0 \leq k_{g} \leq 10$ and $0 \leq k_{r} \leq 10$
            \item $range(I_{100}) \geq 1 MJy\,sr^{-1}$
        \end{enumerate}
    
    For the \textbf{first criterion}, $C_g$ denotes the correlation coefficient between $I_{g}$ and $I_{100}$, and $C_r$ denotes that between $I_{r}$ and $I_{100}$. This criterion aims to ensure the selection of fields with a strong correlation between the optical and IR bands, while considering the presence of cirrus. In the \textbf{second criterion}, $k_g$ represents the linear fitting slope of $I_{g}$ and $I_{100}$, while $k_r$ represents that between $I_{r}$ and $I_{100}$. This criterion excludes fields where we suspect that bright sources have not been adequately removed, as they may significantly impact our linear fit. The \textbf{third criterion} sets a limit on the range of IR intensity within a block to ensure the presence of diffuse and extended structures.

    By applying the three criteria described above, we have selected 2245 blocks out of a total of 24748 ($6187 \times$ 4) blocks, with each block covering an area of 0.25 deg$^2$. The selected cirrus blocks cover an area of 561.25 deg$^2$, which accounts for 9\% of our selected region.

\begin{figure*}[htbp]
        \centering
        \includegraphics[width=\linewidth]{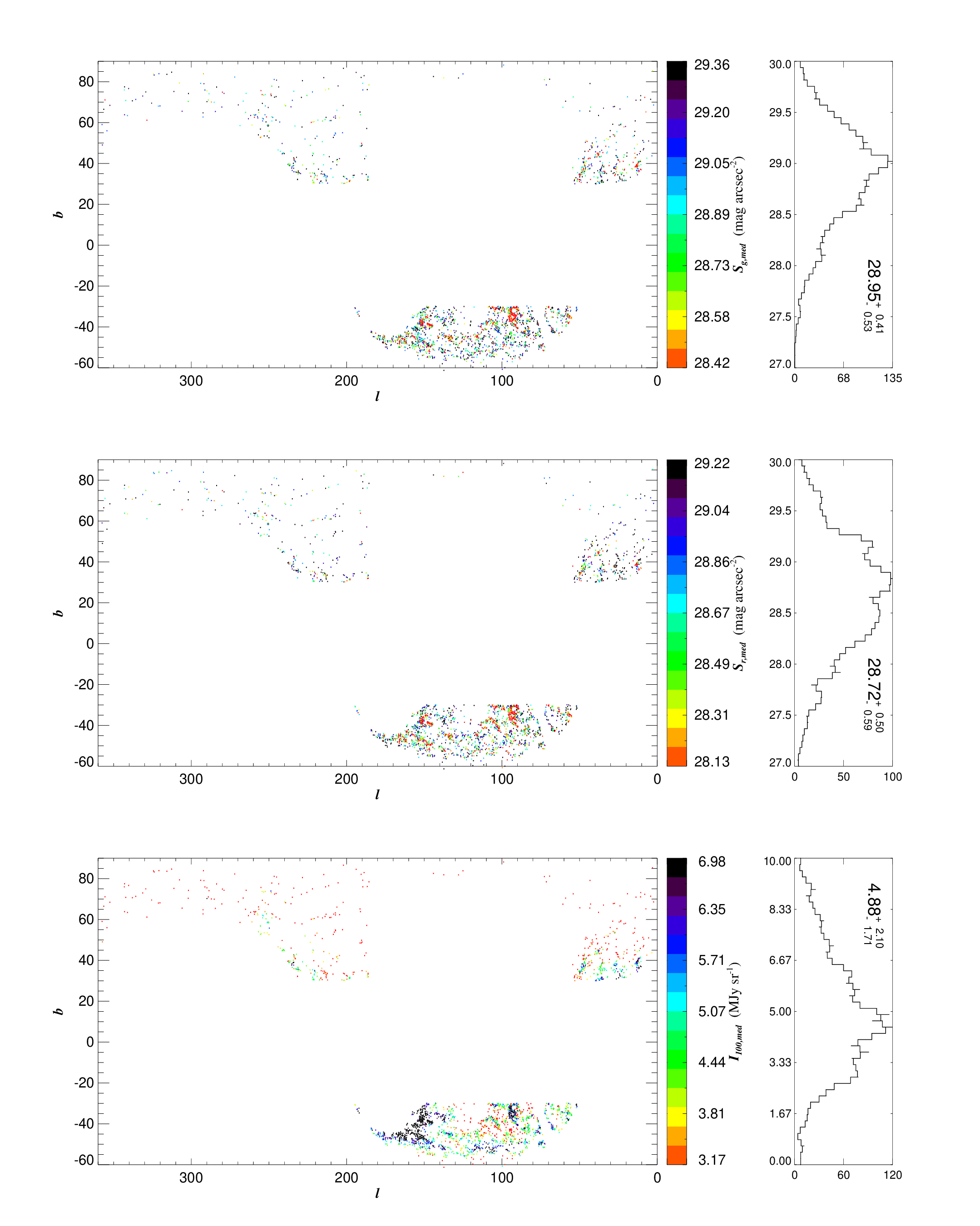}
        \caption{Distributions of $S_{g,med}$, $S_{r,med}$, $I_{100,med}$ in cirrus sample. Each rectangle represents one block covering an area of $0.5\degr \times 0.5\degr$. The color of each rectangle indicates the corresponding value of $S_{g,med}$, $S_{r,med}$, $I_{100,med}$, as illustrated by the colorbars. Histograms and characteristic values are shown on the right side.}
        \label{fig:coverage_all_grid_new_largeplot_intensity}
        \end{figure*}

\begin{figure*}[htbp]
        \centering
        \includegraphics[width=0.9\linewidth]{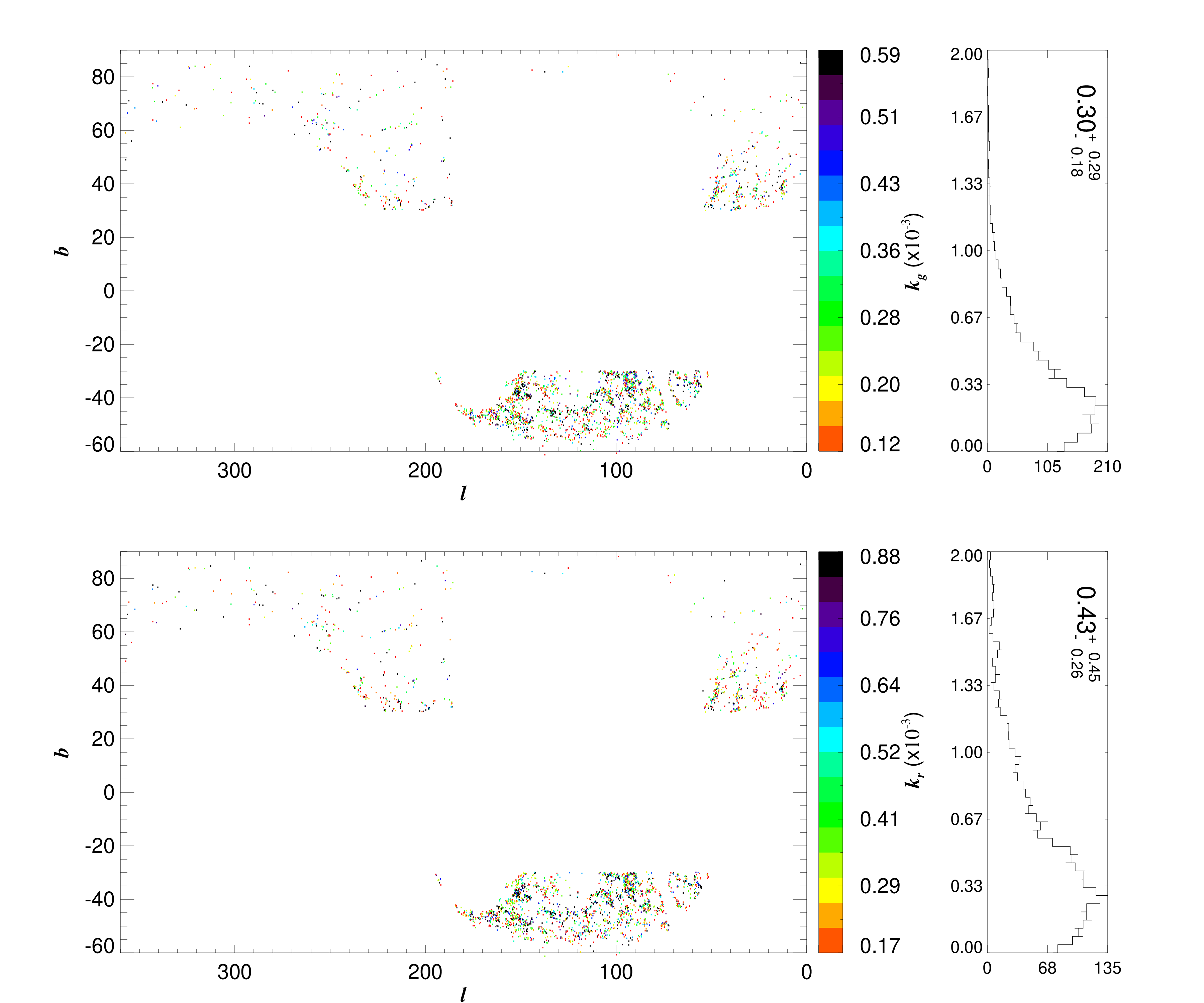}
        \caption{Distributions of $k_g$ and $k_r$ in cirrus sample. Each rectangle represents one block covering an area of $0.5\degr \times 0.5\degr$. The color of each rectangle indicates the corresponding value of $k_g$ and $k_r$, as illustrated by the colorbars. Histograms and characteristic values are shown on the right side.}
        \label{fig:coverage_all_grid_new_largeplot_slope}
        \end{figure*}

    Using these 2245 blocks, we have constructed a comprehensive sample catalog of Galactic cirrus. The catalog includes information such as coordinate, median optical intensity, optical surface brightness, median IR intensity, optical-IR fitting slope, and correlation coefficient for each block. Figure \ref{fig:coverage_all_grid_new_largeplot_intensity} and Figure \ref{fig:coverage_all_grid_new_largeplot_slope} present the distributions of $S_{g,med}$, $S_{r,med}$, $I_{100,med}$, $k_g$ and $k_r$. The optical surface brightness of our cirrus is approximately $28.95^{+0.41}_{-0.53}$  $mag\,arcsec^{-2}$   for the $g$ band and and $28.72^{+0.50}_{-0.59}$  $mag\,arcsec^{-2}$ for the $r$ band. The mean IR intensity is around $4.88^{+2.10}_{-1.71}$  $MJy\,sr^{-1}$. The values $k_g$ and $k_r$ are approximately $0.30^{+0.29}_{-0.18}$ and $0.43^{+0.45}_{-0.26}$, respectively.
    
    We present an example region in Appendix \ref{sec:Appendix_example_region}. The cirrus blocks are outlined with solid boxes, while the non-cirrus blocks are outlined with dotted boxes. Most of the blocks displaying evident cirrus structures have been successfully selected. Additionally, we provide detailed parameters for these blocks in Table \ref{table:cirri information} and \ref{table:cirri parameter}.

\section{Data analysis and results} \label{sec:Data analysis and results}

    \subsection{Optical color and effective temperature}
    
\begin{figure}[htbp]
        \includegraphics[width=0.95\linewidth]{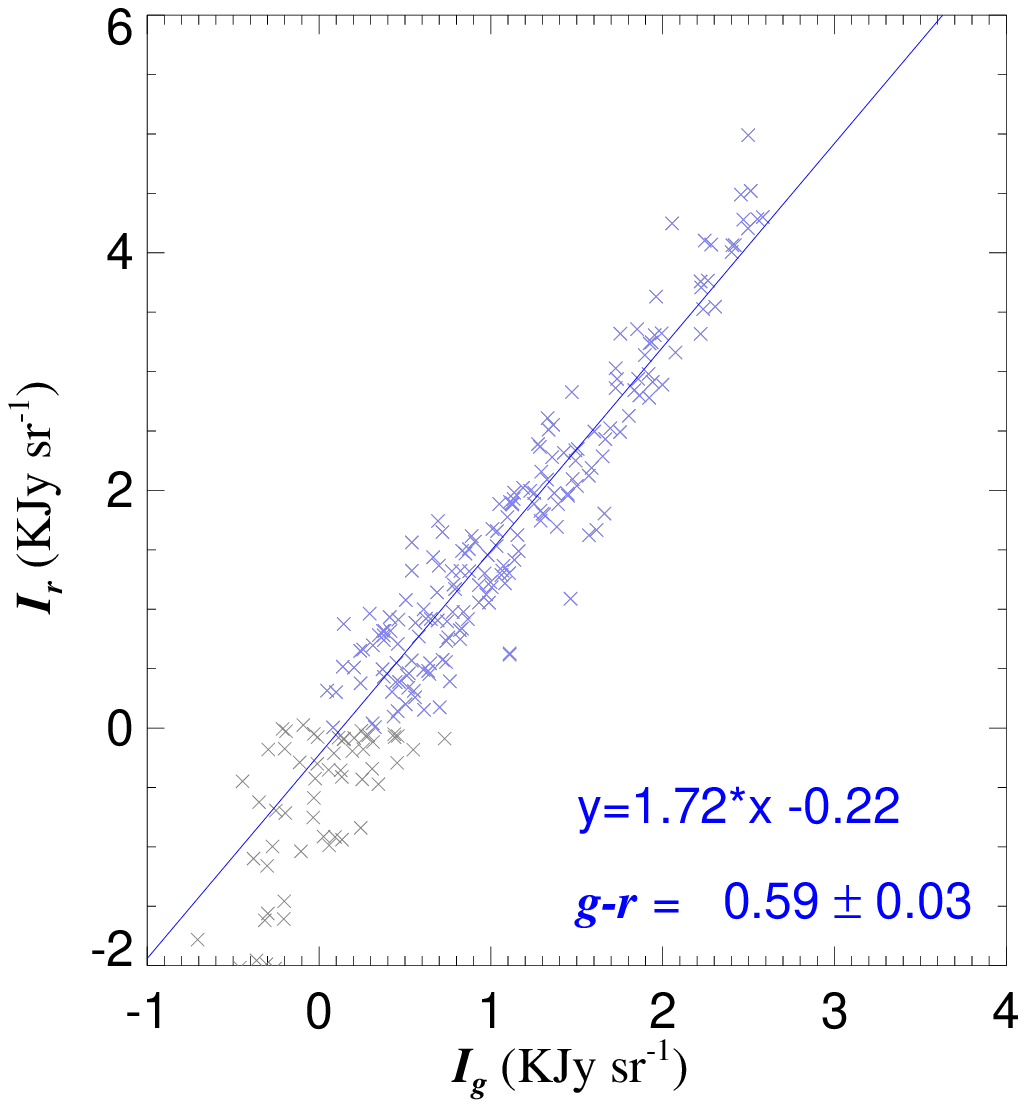}    
        \caption{The linear fitting method for measuring $g-r$ color. The crossed are pixels after mask and iterative remove within one block, the same as that in panel ($c$) of Figure \ref{fig:slope_linear}. The blue crosses are the cirrus pixels and the gray crosses are the background pixels. The blue line is the linear fit result derived from these cirrus pixels. The color g-r can be obtained by converting the fitting slope. }
        \label{fig:color_linear}        
        \end{figure}

\begin{figure*}[htbp]
    \includegraphics[width=0.95\linewidth]{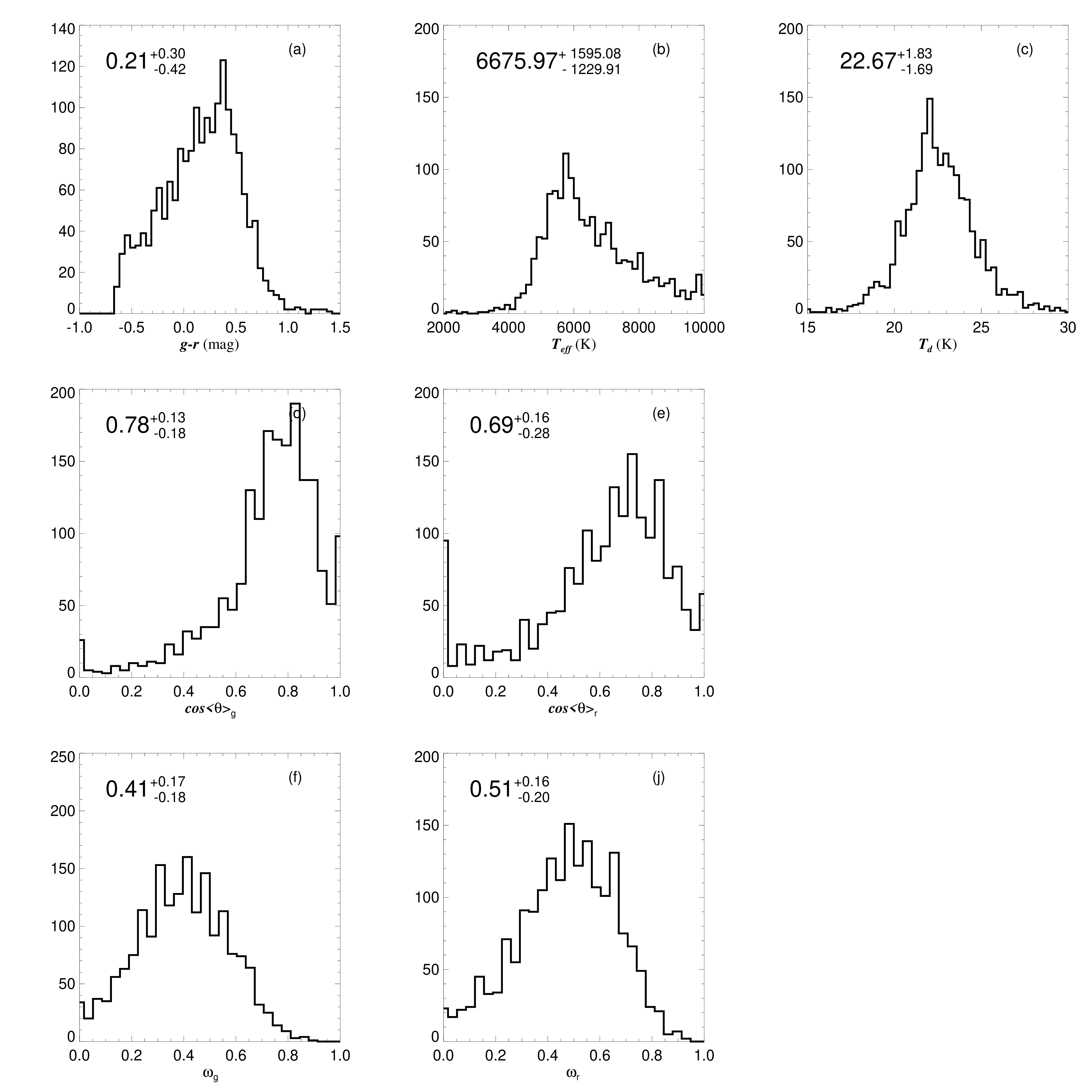}    
    \caption{Histograms of parameters $g-r$, $T_{eff}$, $T_{d}$, $cos \langle \theta \rangle _{g}$, $cos \langle \theta \rangle _{r}$, $\omega_{g}$, and $\omega_{r}$. Median and error values are shown in the upper left corner.}
    \label{fig:parameter_hist}        
    \end{figure*}

    Following a classic color determination method  \citep{guhathakurta_optical_1989,sujatha_galex_2010,murthy_galex_2014,roman_galactic_2020}, we calculate the average $g-r$ color for our cirri. Figure \ref{fig:color_linear} illustrates the method using an example cirrus. For the cirrus pixels as depicted as blue crosses, we perform a linear fitting of the form $I_r = I_g \times m + n$. The fitting slope $m$ can be converted to the $g-r$ color. One advantage of this method is that it accounts for sky background values through the parameter $n$, ensuring that the calculated color is not influenced by the sky background. The histogram of the $g-r$ color is shown in Figure \ref{fig:parameter_hist} ($a$). As evident, the $g-r$ color values are primarily concentrated around $0.21^{+0.30}_{-0.42}$. Previous studies have also calculated the optical color of cirrus. For example, the ring-like low-surface-brightness nebula mentioned in \citet{zhang_mysterious_2020} has a $g-r$ color of 0.3. Other studies by \citet{roman_galactic_2020} and \citet{smirnov_prospects_2023} reported the $g-r$ color range for their cirrus samples as $0.46 \leq g$ $-$ $r \leq 0.93$ and $0.55 \leq g$ $-$ $r \leq 0.73$, respectively. Overall, our cirrus exhibits bluer colors compared to these previous findings.
    The discrepancies mainly arise from the broader spatial coverage of our cirrus observations compared to previous studies. Additionally, the presence of ERE may also affect the color of the cirrus \citep{guhathakurta_optical_1989, witt_extended_2008, ienaka_diffuse_2013, chellew_optical_2022}. This possibility deserves further exploration through the integration of additional optical bands.
    
\begin{figure}[htbp]
        \centering
        \includegraphics[width=0.95\linewidth]{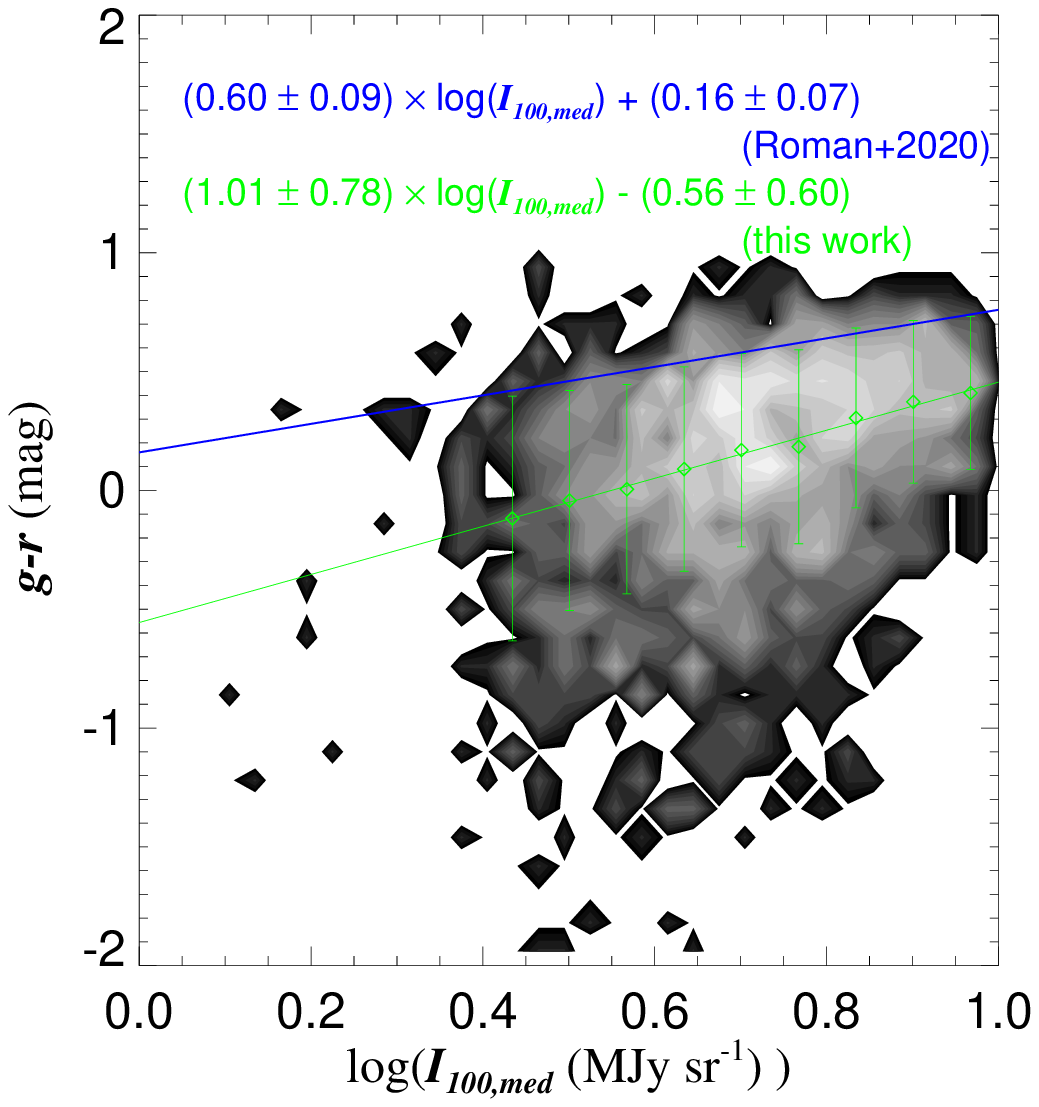}    
        \caption{Contour diagram of pixel density between $g-r$ and $I_{100,med}$ for cirrus sample. The diamonds represent the median values after dividing into bins based on x-axis. The error bars are the uncertainties for $g-r$ in each bin. The green line is our linear fitting result derived from the median values, and the blue line is from \citet{roman_galactic_2020}.}
        \label{fig:color_100mu}        
        \end{figure}

    Then, we compare the $g-r$ color with $I_{100,med}$ as shown in Figure \ref{fig:color_100mu}, revealing a clear correlation. We perform a linear fitting represented by the green line. This correlation indicates that as the 100 $\mu m$ intensity increases, the cirrus tends to have a redder color. This relationship is similar to he findings reported in \citet{roman_galactic_2020}, represented by the blue line in Figure \ref{fig:color_100mu}. There are some differences in the expressions of two lines, which is caused by the difference in the cirrus fields.

    Additionally, we determine the optical effective temperature of our cirrus based on the $g-r$ color. The optical effective temperature is derived through two intensity ratios, utilizing a black body approximation. The temperature can be estimated using the commonly employed $B-V$ color \citep{ballesteros_new_2012}:
        \begin{equation}
        \footnotesize
           T_{eff} = 4600\left(\frac{1}{0.92(B-V)+1.7} + \frac{1}{0.92(B-V)+0.62}\right)
            \label{eq:Teff}
        \end{equation}

    The $B-V$ color is derived from the $g-r$ color using the equation in \citet{jester_sloan_2005} for stars : $B-V$ = 0.98 $\times$ ($g-r$) + 0.22. The panel ($b$) in figure \ref{fig:parameter_hist} shows the histogram of $T_{eff}$. The values of $T_{eff}$ are primarily concentrated around $6675.97^{+1595.08}_{-1229.91}$  K. 
    
%%%%hist parameter%%%%    
%%%%hist parameter%%%%

    \subsection{Dust temperature}
    
    The dust temperature of Galactic cirrus is reported to be approximately 20 K \citep{low_infrared_1984, de_vries_comparison_1985}. For our cirrus clouds, we calculate dust temperatures using the IRAS 60 and 100 $\mu$m intensities. Assuming that the emission at 60 $\mu$m and 100 $\mu$m is optically thin and that the 60 $\mu$m / 100 $\mu$m emission is attributed to blackbody radiation from dust grains at temperature $T_{d}$, we can derive $T_{d}$ using the following expression \citep{schnee_complete_2005}:
        \begin{equation}
            R=0.6^{-(3+\beta)}\frac{e^{144/T_d}-1}{e^{240/T_d}-1} \label{eq:Tdust}
        \end{equation}
    where $R$ represents the ratio of 60  and 100 $\mu m$ intensities, and $\beta$ denotes the emissivity spectral index. The value of $\beta$ depends on various properties of the dust grains, such as composition, size, and compactness. In our analysis, we adopt a value of $\beta$ = 2 and utilize the linear fit slopes of the 60  and 100 $\mu m$ intensities to determine the ratio $R$. By employing this approach, we derive the dust temperatures for all cirrus clouds, and their distribution is presented in the histogram shown in Figure \ref{fig:parameter_hist} ($c$). The dust temperatures of our cirrus clouds are primarily centered around $22.67^{+1.83}_{-1.69}$ K. We note that adopting $\beta$ = 1.5 would result in a $T_{d}$ approximately 1.6 K higher (6.6\% increase) compared to assuming $\beta$ = 2.

%%%%slope vs gbgl and 100
    \begin{figure*}[hbtp]
    \includegraphics[width=\linewidth]{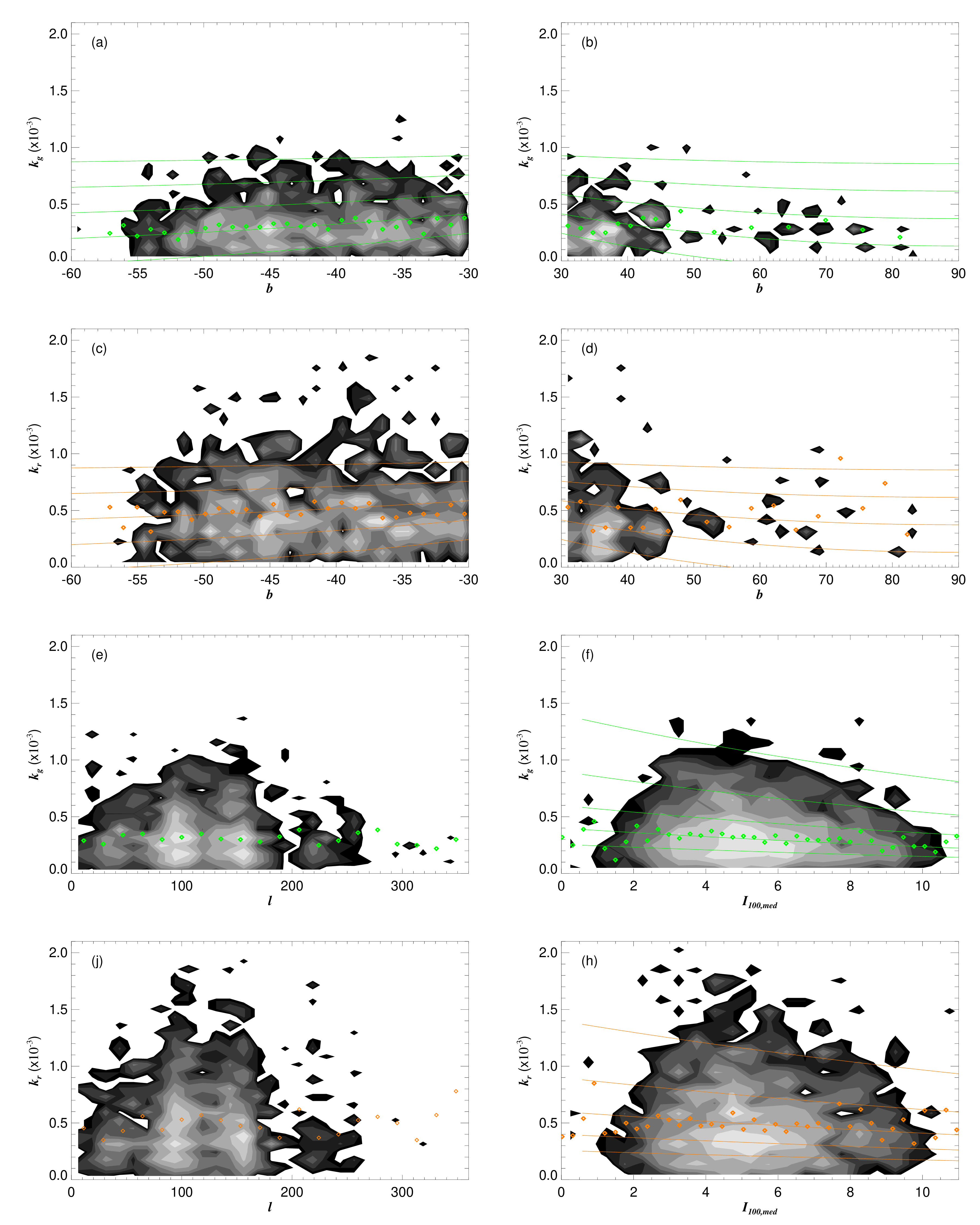}    
    \caption{Correlations between $k_{\lambda}$ and galactic latitude $(b)$, galactic longitude $(l)$, 100 $\mu$m intensity $(I_{100,med})$. Contours of pixel density are plotted. The diamonds are the median values after dividing bins based on x-axis. The green and orange curves in panel ($a$) to ($d$) is the Equation \ref{eq:Jura1979} from \citet{jura_ultraviolet_1979}, from top to bottom with asymmetry factor $cos \langle \theta \rangle$ = 0.20, 0.40, 0.60, 0.80 and 1.00. The green and orange curves in panel $(e)$ and $(f)$ is the Equation \ref{eq:Ienaka2013} from \citet{ienaka_diffuse_2013}, from top to bottom with $\omega$ = 0.70, 0.60, 0.50, 0.40 and 0.30.}
    \label{fig:slope_scatter_gbgl_100}        
    \end{figure*}     
%%%%slope vs gbgl and 100

    \subsection{Dust grain properties}

        \subsubsection{Asymmetry factor}
    
        The forward-scattering effect of dust grains can contribute to variations in scattered light. This effect is commonly evaluated using the asymmetry factor of the phase function, denoted as $cos \langle \theta \rangle$. When there is strong forward scattering, $cos \langle \theta \rangle$  tends to be close to 1, whereas for isotropic scattering, $cos \langle \theta \rangle$ is around 0. Taking into account the scattering anisotropy within the diffuse interstellar medium (ISM), \citet{jura_ultraviolet_1979} proposed a model for scattered starlight as a function of $\left| b \right|$. In this model, the assumption is made that the clouds are illuminated by an infinite and homogeneous plane. The ratio of scattered light intensity to interstellar 100 $\mu$m intensity can be approximately determined as follows \citep{jura_ultraviolet_1979}:
            \begin{equation}
                I_{\lambda,sca} \propto 1-1.1cos \langle \theta \rangle(sin \left| b \right|)^{1/2} \label{eq:Jura1979}
            \end{equation}

        We calculate the asymmetry factors for all cirrus clouds in the $g$ and $r$ bands, and the histograms are shown in Figure \ref{fig:parameter_hist} ($d$) and ($e$). From the figures, we can see that $cos \langle \theta \rangle$ in the $g$ band is primarily concentrated around $0.78^{+0.13}_{-0.18}$. In the $r$ band, $cos \langle \theta \rangle$ is centered around $0.69^{+0.16}_{-0.28}$. This discrepancy suggests that the forward-scattering characteristic is more pronounced in the $g$ band compared to the $r$ band.

        We compare the one-dimensional distribution of $k_{\lambda}$ with Galactic latitudes  and Galactic longitudes, respectively, in Figure \ref{fig:slope_scatter_gbgl_100}. In general, the values of $k_{\lambda}$ exhibit a decreasing trend with increasing $|b|$, while no clear trend is observed with $l$. The dependence of DGL on Galactic latitudes has been identified in previous studies \citep{sano_first_2016, sano_galactic_2017}. We plot the decreasing relationship using Equation \ref{eq:Jura1979} from \citet{jura_ultraviolet_1979} as green curves in Figure \ref{fig:slope_scatter_gbgl_100}. However, we find that the decreasing curve obtained using a single asymmetry factor fails to fully account for the scatter of slopes. To address this, we incorporate the forward-scattering effect by plotting Equation \ref{eq:Jura1979} with different values of $cos \langle \theta \rangle$ (0.20, 0.40, 0.60, 0.80, and 1.00) from top to bottom. By considering the impact of various asymmetry factors, we successfully explain the large scatter of slopes.

        \subsubsection{Albedo}
        
        Albedo reflects the proportion of incident radiation that is scattered by the dust. \citet{ienaka_diffuse_2013} proposed a unified model suggesting  that the slope $k_{\lambda}$ is influenced by both the optical depth and the albedo. In this model, it is assumed that a plane-parallel dust slab is illuminated by monochromatic starlight from the back. The model can be mathematically expressed as follows \citep{ienaka_diffuse_2013}:
            \begin{equation}
            \footnotesize
                k_{\lambda} \propto exp[-(1-\omega)\tau][1-exp(-\omega\tau)]/\{1-exp[-(1-\omega)\tau]\} \label{eq:Ienaka2013}
            \end{equation}  
        where $\omega$ is albedo and $\tau$ is optical depth. Provided that $I_{100}/A_{V}$ = 10 $MJy\,sr^{-1}\,mag^{-1}$ and $R_{V}=3.1$, we can calculate the albedo values $\omega$ for all cirrus clouds in the $g$ and $r$ bands. The histograms depicting these calculations are shown in Figure \ref{fig:parameter_hist} ($f$) and ($j$). The calculated albedo values are found to be concentrated around $0.41^{+0.17}_{-0.18}$ in the $g$ band and $0.51^{+0.16}_{-0.20}$ in the $r$ band. It is found that the albedo values are higher in the $r$ band, indicating a greater amount of light scattered in the $r$ band compared to that in the $g$ band.

        Similarly, we analyze the relationship between $k_{\lambda}$ and $I_{100,med}$ in Figure \ref{fig:slope_scatter_gbgl_100}. In this figure, it is evident that there is a decreasing trend in $k_{\lambda}$ as $I_{100,med}$ increases. Initially, we plot Equation \ref{eq:Ienaka2013} using a single albedo value, which provides a satisfactory description of the decreasing relationship. However, this curve with a single albedo value fails to capture the significant up-and-down scatter observed in the slopes. To address this issue, we further incorporate Equation \ref{eq:Ienaka2013} with varying albedo values, specifically $\omega$ = 0.70, 0.60, 0.50, 0.40, and 0.30 from top to bottom. By doing so, we are able to effectively explain the substantial scatter observed in the slopes, thus providing a more comprehensive understanding of the data.
        
        In summary, this section focuses on determining the dust grain properties of cirrus clouds, specifically the asymmetry factor $cos \langle \theta \rangle$ and albedo $\omega$. Our analysis has revealed that the variation of slope $k_{\lambda}$ with $b$ and $I_{100,med}$ can explain part of the scatter. However, the main cause of scatter is attributed to the differences in dust grains properties within the cirrus.

\section{Discussion} \label{sec:Discussion}

    \subsection{$k_{g}$ vs $k_{r}$}

\begin{figure*}[htbp]
        \centering
        \includegraphics[width=0.88\linewidth]{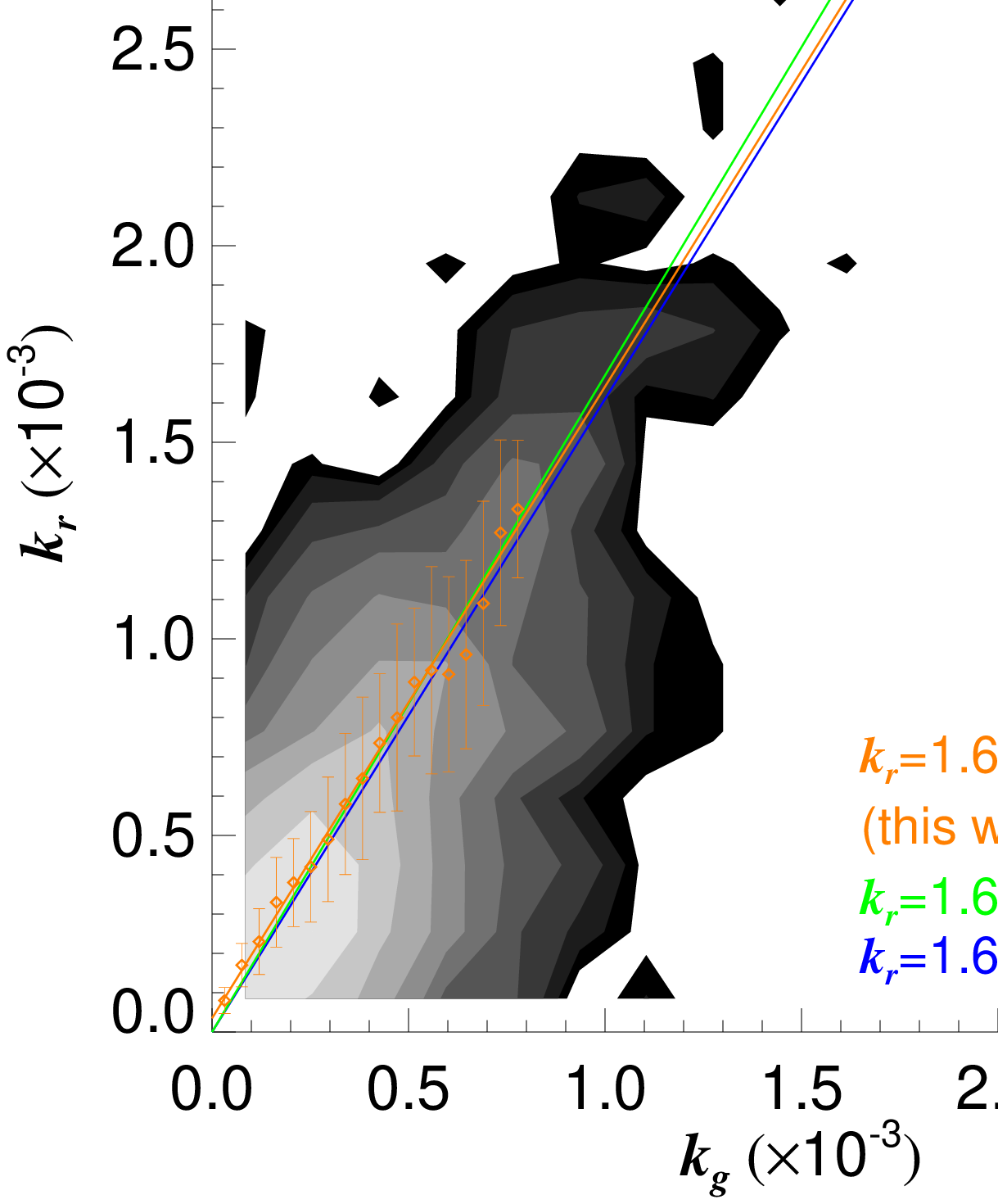}    
        \caption{Left panel: contour plot of $k_g$ and $k_r$. The orange points are median values after dividing into bins based on x-axis. The orange line is the linear fitting result derived from the median values. And the blue and green lines are indicated by the stars in the left panel. Right panel: optical spectra of DGL from \citet{brandt_spectrum_2012}. The orange stars are median values from our work correspond to the g band and r band respectively. They have been scaled with a biased factor of 2.70 for comparison with others stars. The blue and green stars represent the $k_{\lambda}$ at the g and r bands.}
        \label{fig:slope_hist}        
        \end{figure*}

    We compare the slopes $k_{g}$ and $k_{r}$ for all cirrus clouds, as illustrated in the left panel of Figure \ref{fig:slope_hist}. It is evident from the figure that there are significant differences in the slopes between the $g$ and $r$ bands. Overall, the slopes in the $r$ band  are larger than those in the $g$ band. This finding suggests that, given the same incident intensity, a greater amount of starlight is scattered in the $r$ band.  The left panel of Figure \ref{fig:slope_hist} presents the histograms of the slopes, with $k_g$ being concentrated around $0.30^{+0.29}_{-0.18}$ and $k_{r}$ being concentrated around $0.43^{+0.45}_{-0.26}$. To quantify the relationship between $k_g$ and $k_r$, we fit a linear model represented by the orange line. The fitting equation is expressed as $k_r$ = 1.61 $\times$ $k_g$ + 0.03.
     
    The optical spectrum of DGL was measured by \citet{brandt_spectrum_2012} using 92,000 blank sky spectra. The results obtained were in agreement with scattered starlight, exhibiting characteristics such as a clear 4000 $\rm \AA$ break and broad Mgb absorption. This behavior could be accurately replicated using a simple radiative transfer model of the Galaxy. In the right panel of Figure \ref{fig:slope_hist}, we present the DGL spectra from different models, and we obtain the the ratio of $k_r$ to $k_g$, which is 1.61 or 1.67, as shown by the blue and green stars. We also plot our median values as orange stars. Notably, our fitting outcomes align well with these ratio in the left panel, as shown by the blue and green lines. This agreement implies the potential illumination of our cirrus clouds by the ISRF.

    \subsection{Large scatter of slope}

        \subsubsection{Scatter in our large sample}
        
        As found in previous studies, there is a significant scatter of slopes observed across different fields, with variations spanning a range of 4--5 times. In our study, we have constructed a substantial sample of cirrus clouds in the high Galactic latitudes region and determined their respective slopes. Similarly, we have observed a substantial scatter within our sample, as shown in Figure \ref{fig:slope_hist}. Subsequently, we will discuss the factors that contribute to this significant scatter in slopes.

        \subsubsection{Slope vs color and $T_d$}
        
\begin{figure*}[htbp]
            \centering
            \includegraphics[width=0.63\linewidth]{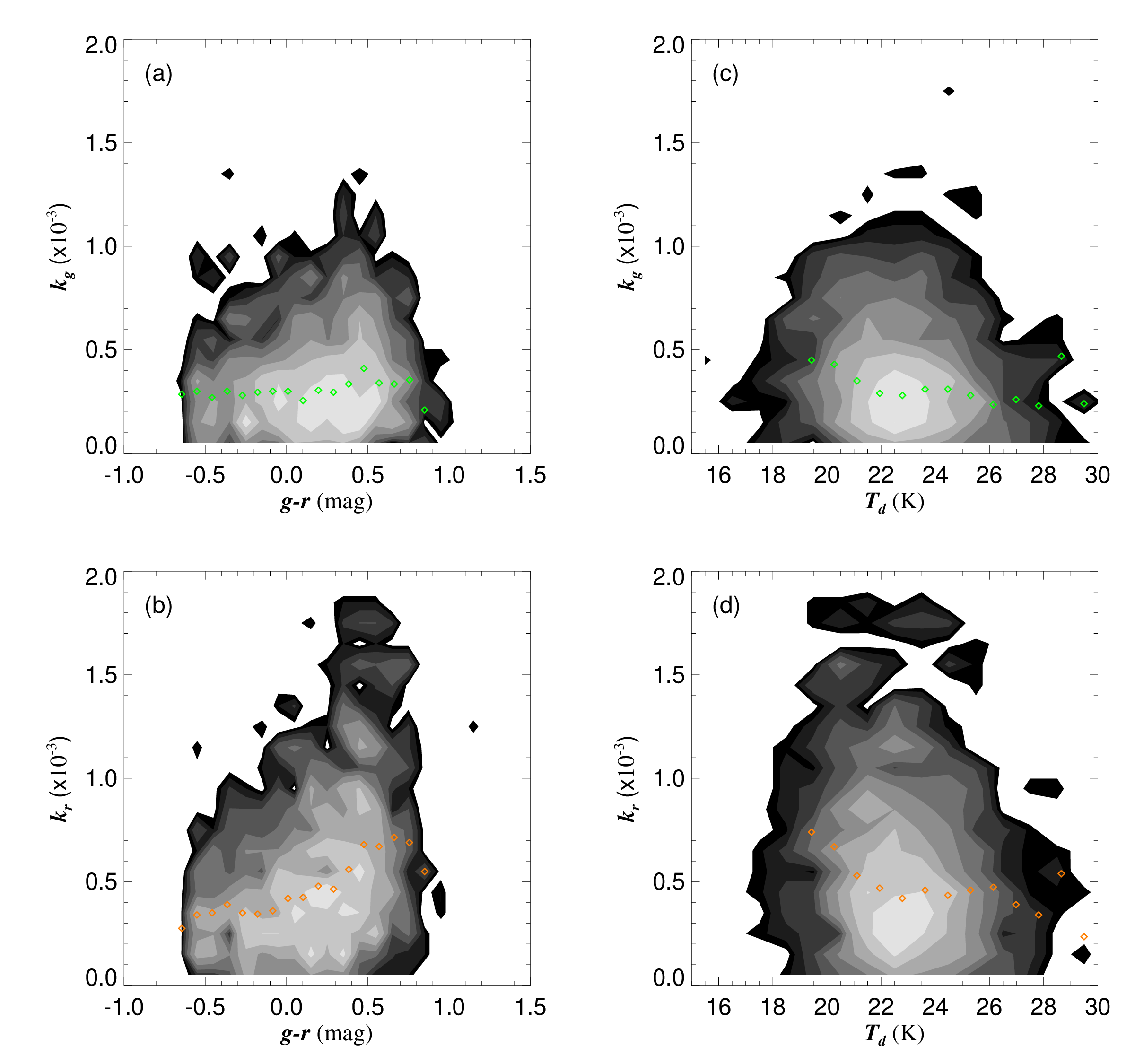}
            \caption{Correlations between slope and color $g-r$, dust temperature $T_{d}$. The diamonds represent the median points after dividing into bins based on x-axis. }
            \label{fig:slope_scatter_color_T}        
            \end{figure*}

        In Section \ref{sec:Data analysis and results}, we find that the variation of slopes with factors such as $b$ and $I_{100}$ can account for a portion of the observed scatter. However, it is evident that the properties of the dust grains contribute significantly to the overall scatter of slopes. In addition, we investigate the relationship between slopes and other variables, including $g-r$ colors and dust temperature $T_d$. However, upon careful examining Figure \ref{fig:slope_scatter_color_T}, we did not find any apparent correlations.
        Note that color and dust temperature serve as indicators of the impact of the ISRF on cirrus characteristics. While the absence of obvious trends can be interpreted as the ISRF not being the main cause of the scatter in slopes.
        
        \subsubsection{Asymmetry factor and albedo}
        
        In the previous sections, we have determined that dust grain properties, specifically the asymmetry factor and albedo, are the primary contributors to the significant scatter observed in slopes. Moving forward, we will now discuss the spatial distribution of these asymmetry factors and albedo values. 
        
            \begin{figure*}[htbp]
            \centering
            \includegraphics[width=\linewidth]{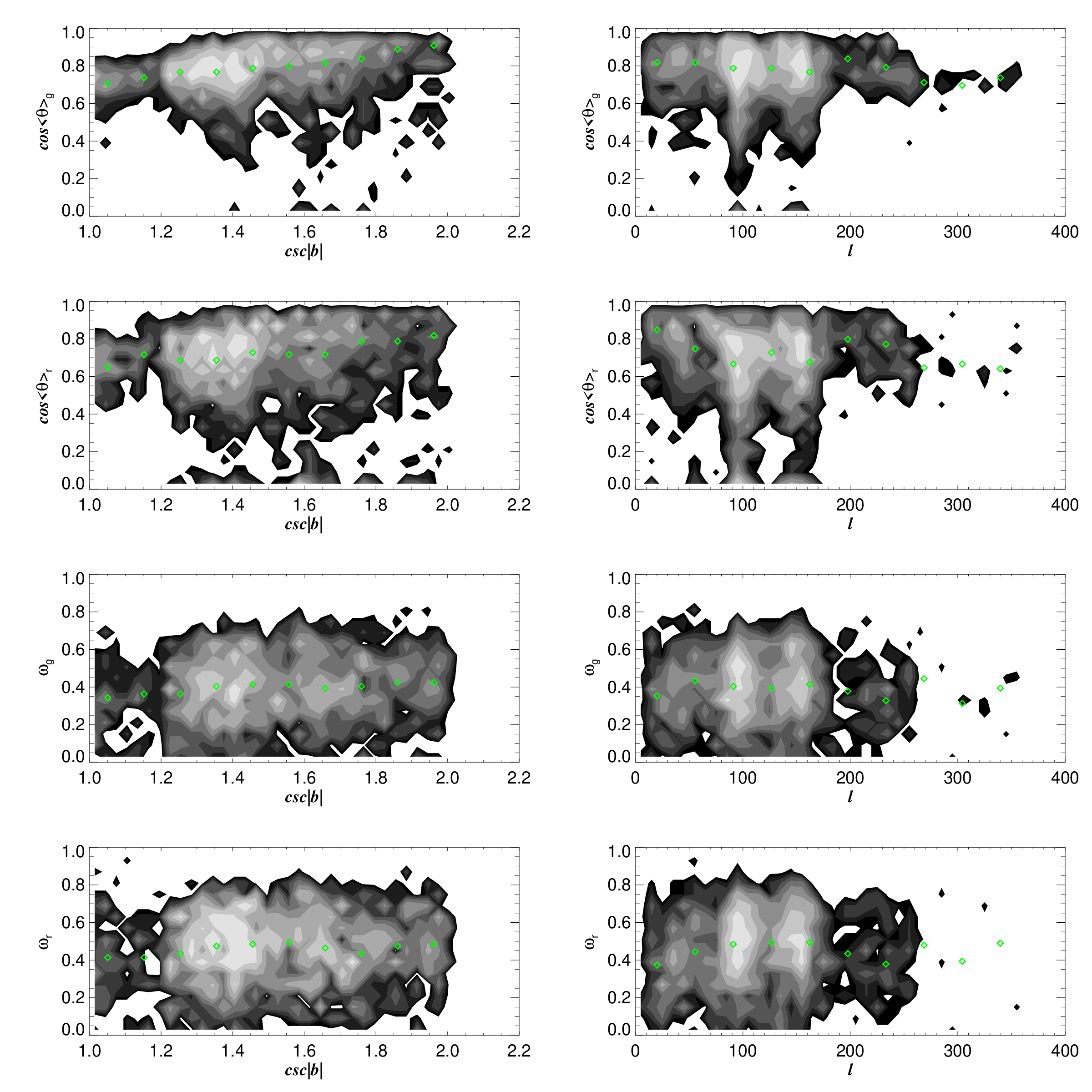}
            \caption{Spatial distribution of asymmetry factor and albedo with b and l. The diamonds represent the median points after dividing into bins based on x-axis. In left columns, $csc\lvert b \rvert$ is used instead of b. }
            \label{fig:asym_albedo_gbgl}        
            \end{figure*}  
            
        Figure \ref{fig:asym_albedo_gbgl} exhibits the one-dimensional distribution of asymmetry factors and albedo in relation to $b$ and $l$. Within this figure, it is noticeable that the values of the asymmetry factor and albedo exhibit no significant dependence on $l$ or $b$. This suggests that changes in the dust properties are localized, with no significant large-scale gradient variations.

\begin{figure}[htbp]
            \centering
            \includegraphics[width=0.83\linewidth]{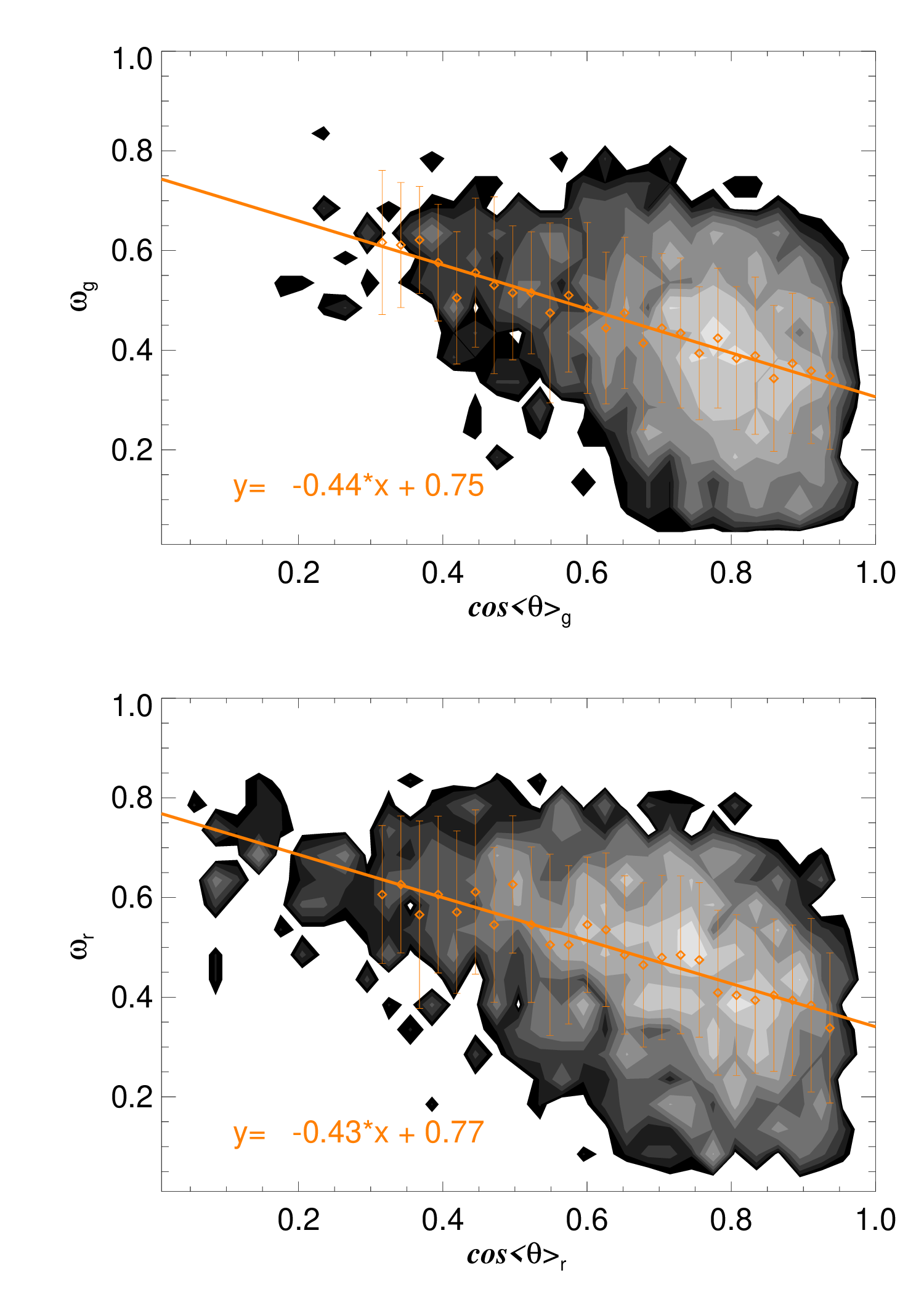}
            \caption{Comparison between asymmetry factor and albedo. The diamonds represent the median points after dividing into bins based on x-axis. And the orange lines are the linear fitting results derived from the median values.}
            \label{fig:asym_albedo}        
            \end{figure}

        Additionally, we examine the relationship between the asymmetry factor and albedo for our cirrus clouds in the $g$ and $r$ bands in Figure \ref{fig:asym_albedo}. We observe an anti-correlation trend between the asymmetry factor and albedo. This trend suggests that the ratio of scattered light to incident light is influenced, in part, by whether the light is scattered straight ahead or at an angle. In the case of strong forward scattering, when $cos \langle \theta \rangle$ $\sim$ 1, less light is scattered due to the scattered light's direction not necessarily aligning with our line of sight. 
        
\section{Conclusion} \label{sec:Conclusion}

    In this paper, we employ the optical-IR correlation to identify Galactic cirrus clouds and obtain their corresponding parameters. Subsequently, we delve into an analysis of the factors contributing to the significant scatter in slopes, including the spatial distribution, external radiation field, and dust properties. Our main conclusions are as follows:

    \begin{enumerate}
        \item 
        We have successfully identified Galactic cirrus clouds across a total area of 6,187 $deg^{2}$ in high Galactic latitudes. From this extensive survey, we have selected  561.25 $deg^{2}$ fields that exhibit distinct cirrus structures. As a result, we have constructed a large sample of cirrus clouds situated at high Galactic latitudes ($|b| \geq 30\degr$). The optical surface brightness of our identified cirrus clouds measures around $28.95^{+0.41}_{-0.53}$  $mag\,arcsec^{-2}$ in the $g$ band and $28.72^{+0.50}_{-0.59}$   $mag\,arcsec^{-2}$ in the $r$ band. Additionally, the average infrared intensity of these cirrus clouds is approximately $4.88^{+2.10}_{-1.71}$ $MJy\,sr^{-1}$.
        \item
        The comparison between $k_g$ and $k_r$ aligns with the DGL model of scattered light, as proposed by \citet{brandt_spectrum_2012}. This result strongly indicates that our identified cirrus clouds are indeed illuminated by the diffuse, scattered starlight originating from ISRF.
        \item
        We calculate some parameters of cirri. The optical color $g-r$ is concentrated around $0.21^{+0.30}_{-0.42}$. The optical effective temperature $T_{eff}$ is around $6675.97^{+1595.08}_{-1229.91}$ K. The dust temperature  $T_d$ is around $22.67^{+1.83}_{-1.69}$ K. The asymmetry factor $cos<\theta>$ is around $0.78^{+0.13}_{-0.18}$ and $0.69^{+0.16}_{-0.28}$ of $g$ and $r$ bands. The albedo $\omega$ is around $0.41^{+0.17}_{-0.18}$ and $0.51^{+0.16}_{-0.20}$ of $g$ and $r$ bands. 
        \item
        We have observed a significant scatter in slopes within our large sample, prompting us to investigate the underlying causes for this phenomenon. While we found that slopes decrease as $b$ and $I_{100,med}$ decrease, we concluded that these factors alone are insufficient in explaining the observed large scatter in slopes. Upon further examination of various dust properties, such as the asymmetry factor and albedo, we discovered that they play a crucial role in explaining the substantial scatter in slopes. In contrast, we did not find any correlation between slopes and color or dust temperature, indicating that variations in the external radiation field are inadequate in accounting for the observed scatter in slopes. Therefore, we conclude that the primary driver of the scatter in slopes is the variations in dust properties.
        \item 
        The asymmetry factor and albedo are independent of either $l$ or $b$. This indicates that the variations in dust properties occur at a localized level and do not exhibit significant large-scale gradients.
    \end{enumerate}

\begin{acknowledgments}

The authors appreciate the anonymous referee for the valuable comments and suggestions that improved the manuscript significantly. W.Z. thanks Hu Zou for helpful suggestions. This work is supported by the National Natural Science Foundation of China (NSFC) (No. 12090041; 12090040) and the National Key R\&D Program of China grant (No. 2021YFA1600401; 2021YFA1600400). We acknowledge the science research grants from the China Manned Space Project with NO. CMS-CSST-2021-B06. This work is also sponsored by NSFC (No. 11733006) and the Strategic Priority Research Program of the Chinese Academy of Sciences (No. XDB0550100). 

\end{acknowledgments}

\clearpage
\appendix

     \section{An example region} \label{sec:Appendix_example_region}
    
    In this section, we present four images of an example region: the residual $g$-band image from DESI, the masked $g$-band image, the convolved and rebinned $g$-band image, and the IRAS 100 $\mu m$ image. The example region consists of 17 fields that have been seamlessly stitched together, resulting in a field of view (FOV) of $6\fdg3 \times 3\fdg5$. To aid in visual clarity, we have delineated the 29 cirrus blocks using solid lines, while the no-cirrus blocks are outlined with dotted lines. It is evident that the majority of the blocks exhibiting evident cirrus structures have been accurately identified and selected. Furthermore, a comprehensive list of detailed parameters for these identified blocks has been provided in Table \ref{table:cirri information} and \ref{table:cirri parameter}.

        \begin{figure}[htbp]
        \centering
        \includegraphics[width=0.85\textwidth]{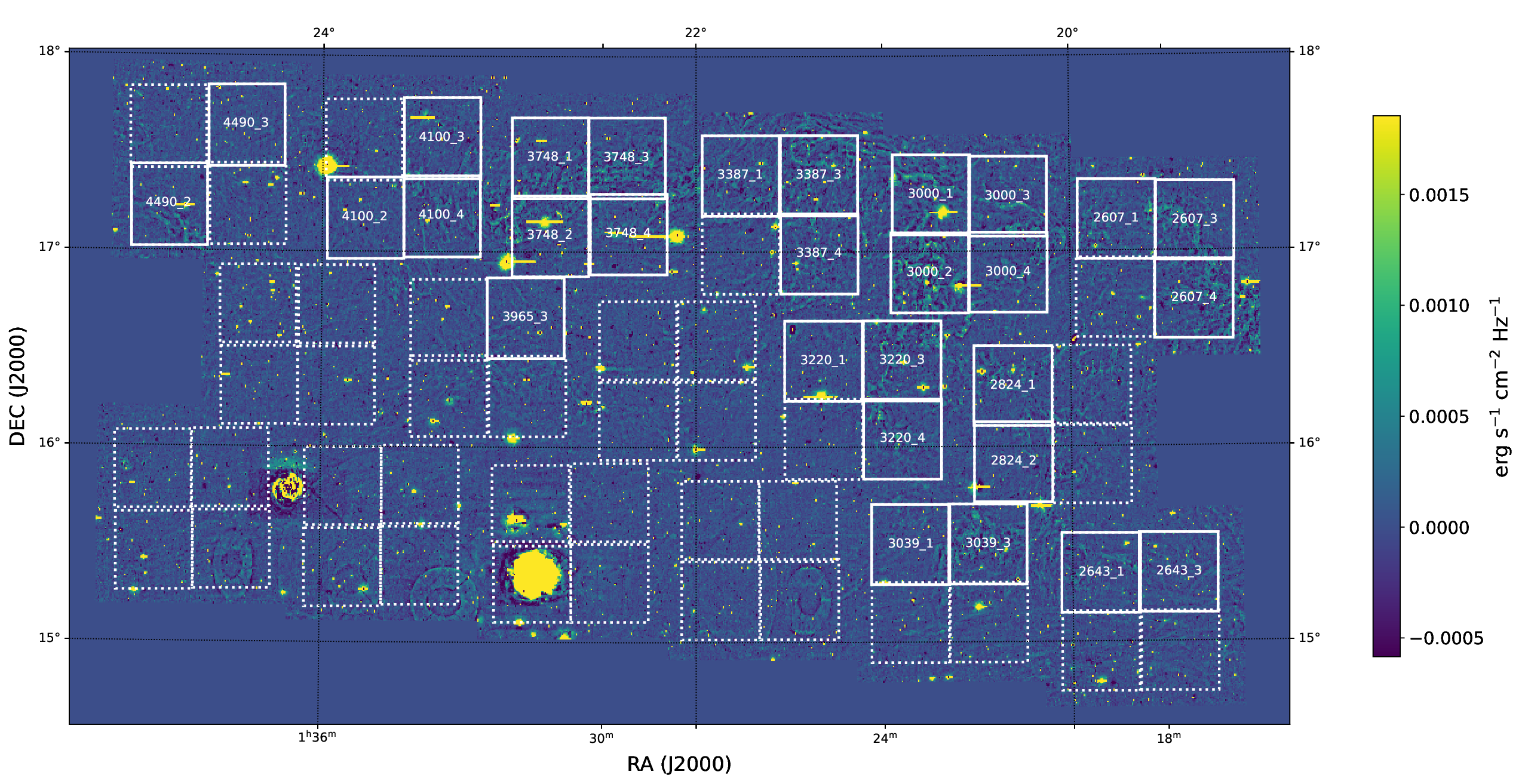}
        \end{figure}

        \begin{figure}[htbp]
        \centering
        \includegraphics[width=0.85\textwidth]{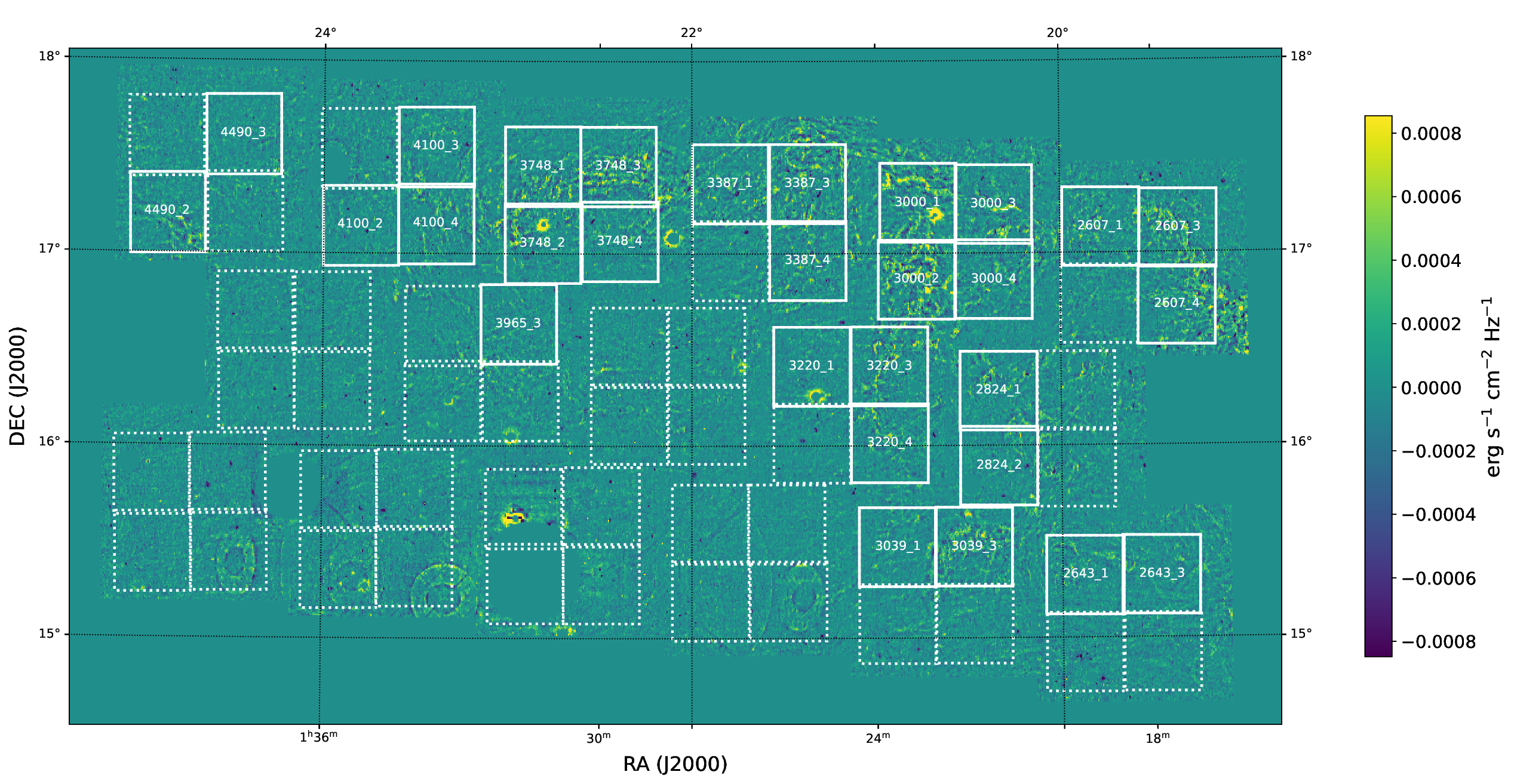} 
        \end{figure}   
        
        \begin{figure}[htbp]
        \centering
        \includegraphics[width=0.85\textwidth]{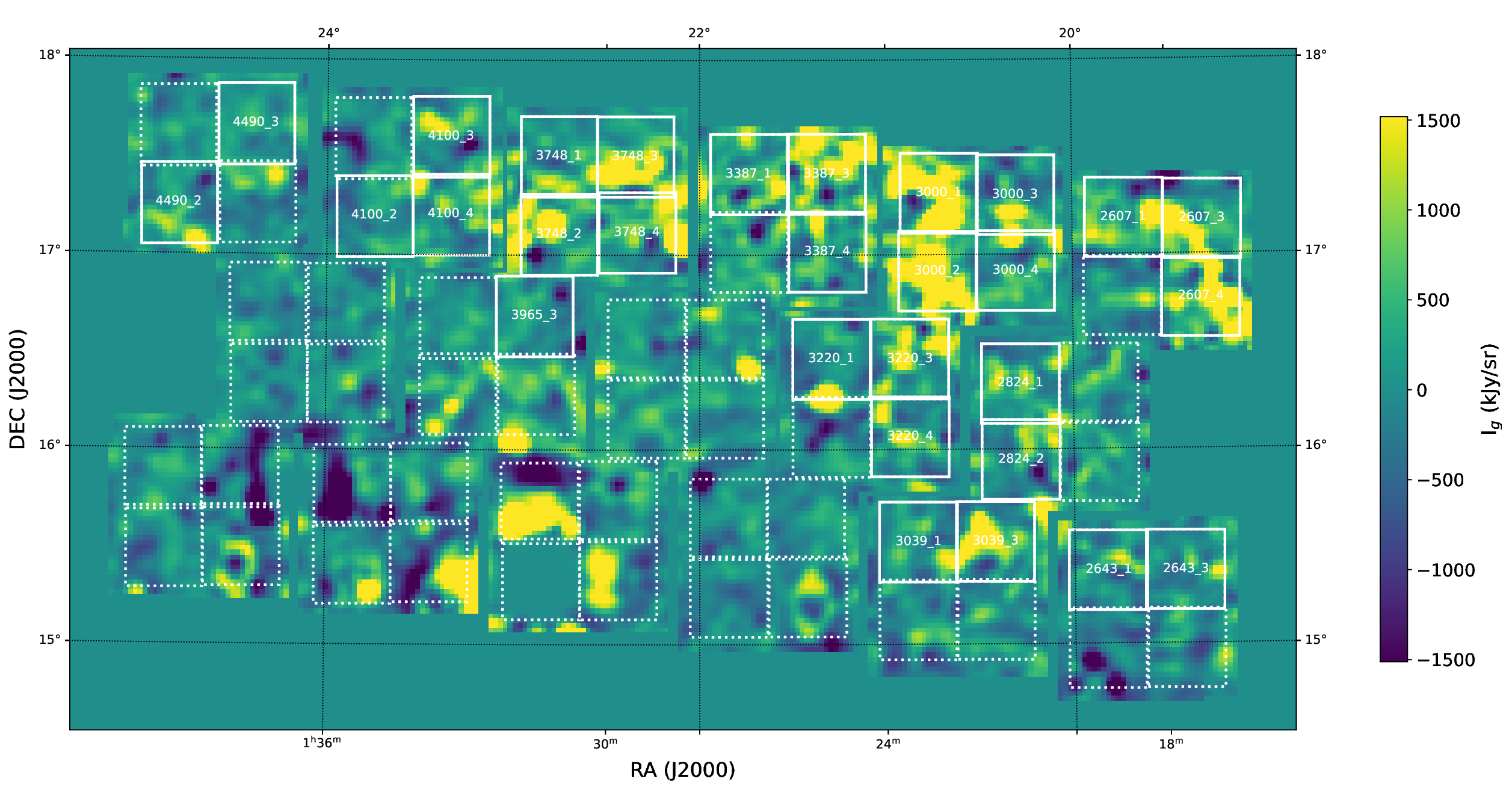} 
        \end{figure}     
        
        \begin{figure}[htbp]
        \centering
        \includegraphics[width=0.85\textwidth]{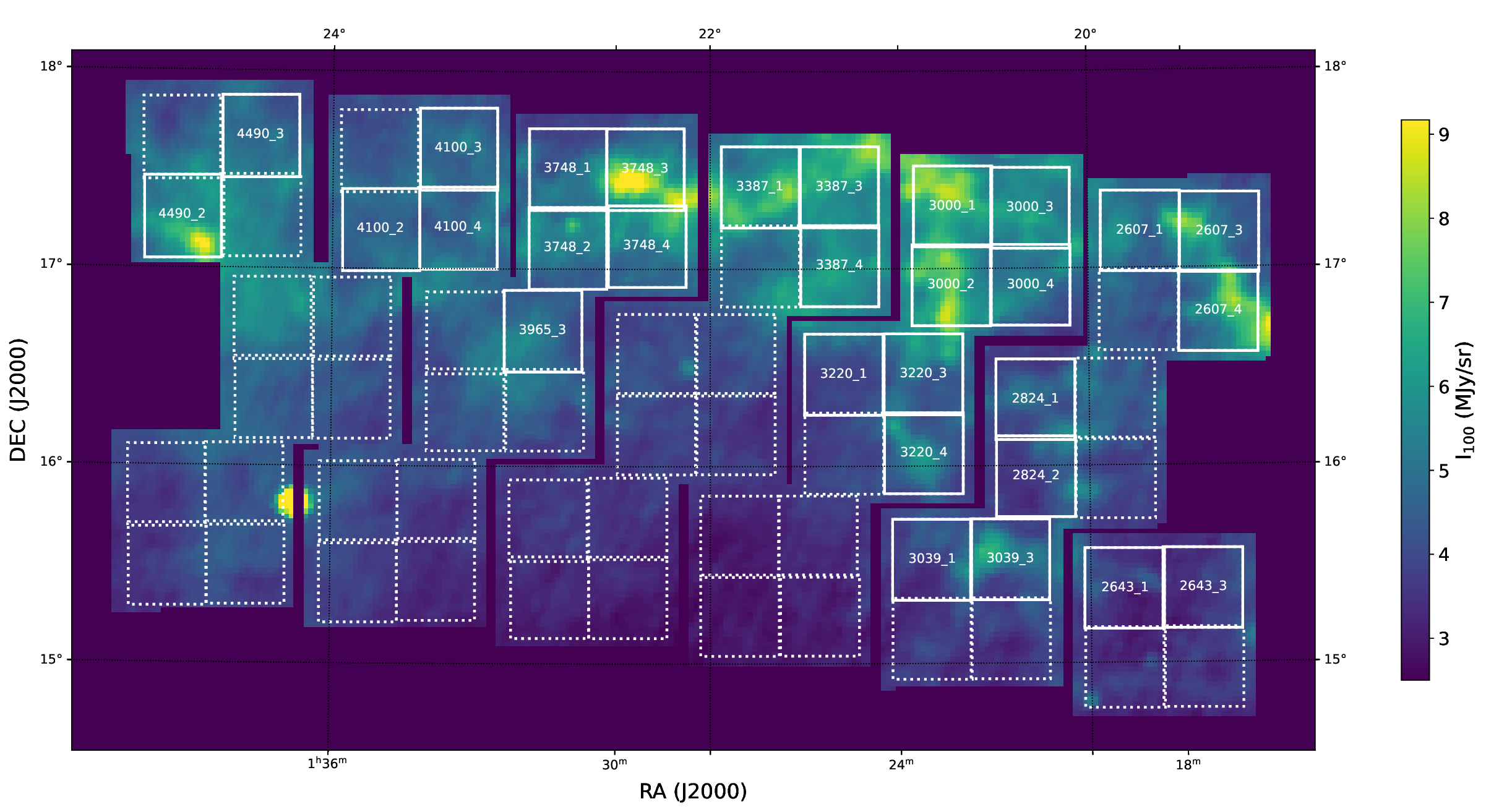}       
        \caption{Example region with a FOV of $6.3^\circ \times 3.5^\circ$, which is made up of 17 fields stitched. From top to bottom: DESI residual $g$-band image, masked $g$-band image, convolved and rebinned $g$-band image, and IRAS 100 $\mu m$ image. The solid frames are cirrus blocks and the dotted frames are no-cirri blocks. The numbers are field IDs.}
        \label{fig:plotbox_3images}        
        \end{figure}

    \section{Distributions of parameters calculated} \label{sec:Distributions of parameters calculated}
    
\begin{figure*}[htbp]
        \centering
        \includegraphics[width=0.9\linewidth]{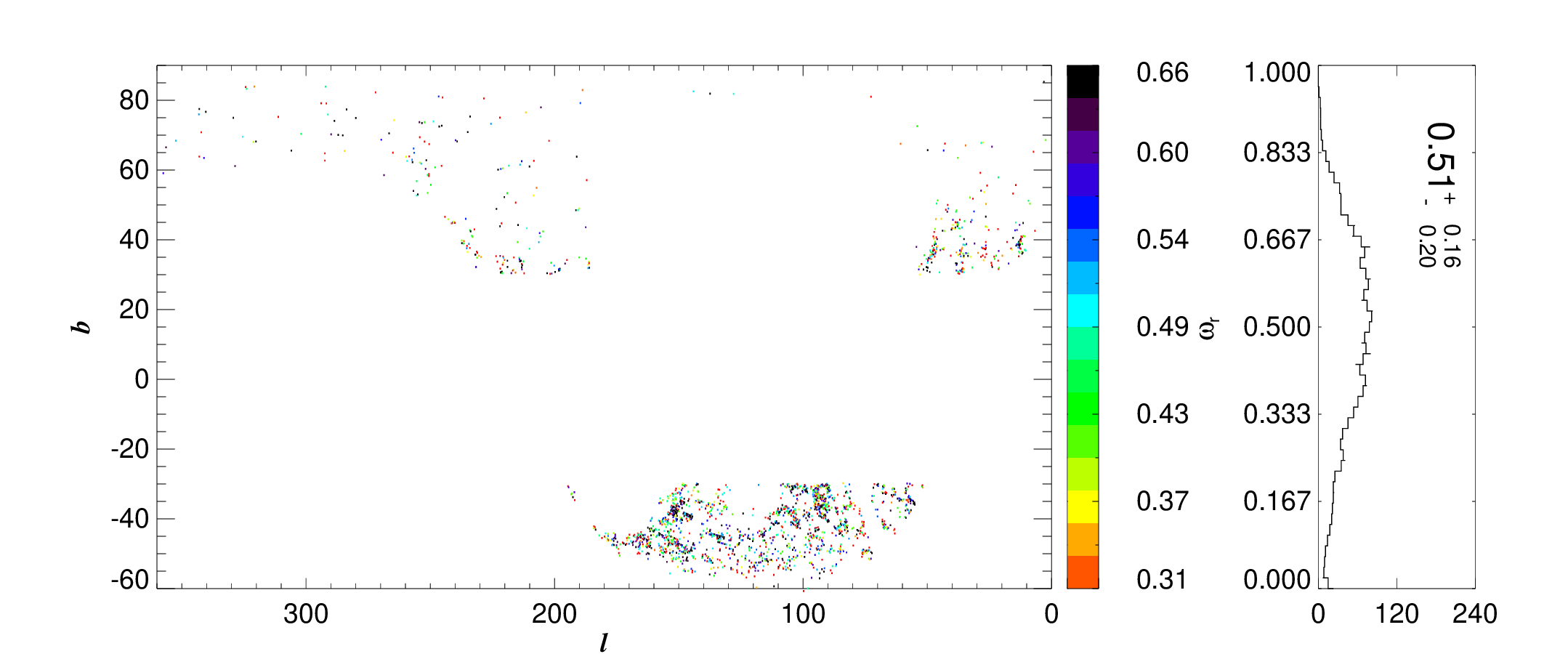}
        \caption{Distributions of g-r, $T_{eff}$, $T_d$, $cos<\theta>_g$, $cos<\theta>_r$, $\omega_g$, $\omega_r$. Same as in the Figure \ref{fig:coverage_all_grid_new_largeplot_intensity} and \ref{fig:coverage_all_grid_new_largeplot_slope}, but for parameters. }
        \label{fig:coverage_para_all_grid_albedo_r}        
        \end{figure*}

    Figure \ref{fig:coverage_para_all_grid_albedo_r} shows the distributions of parameters calculated above.
    
        \begin{figure*}[htbp]
        \centering
        \includegraphics[width=0.9\linewidth]{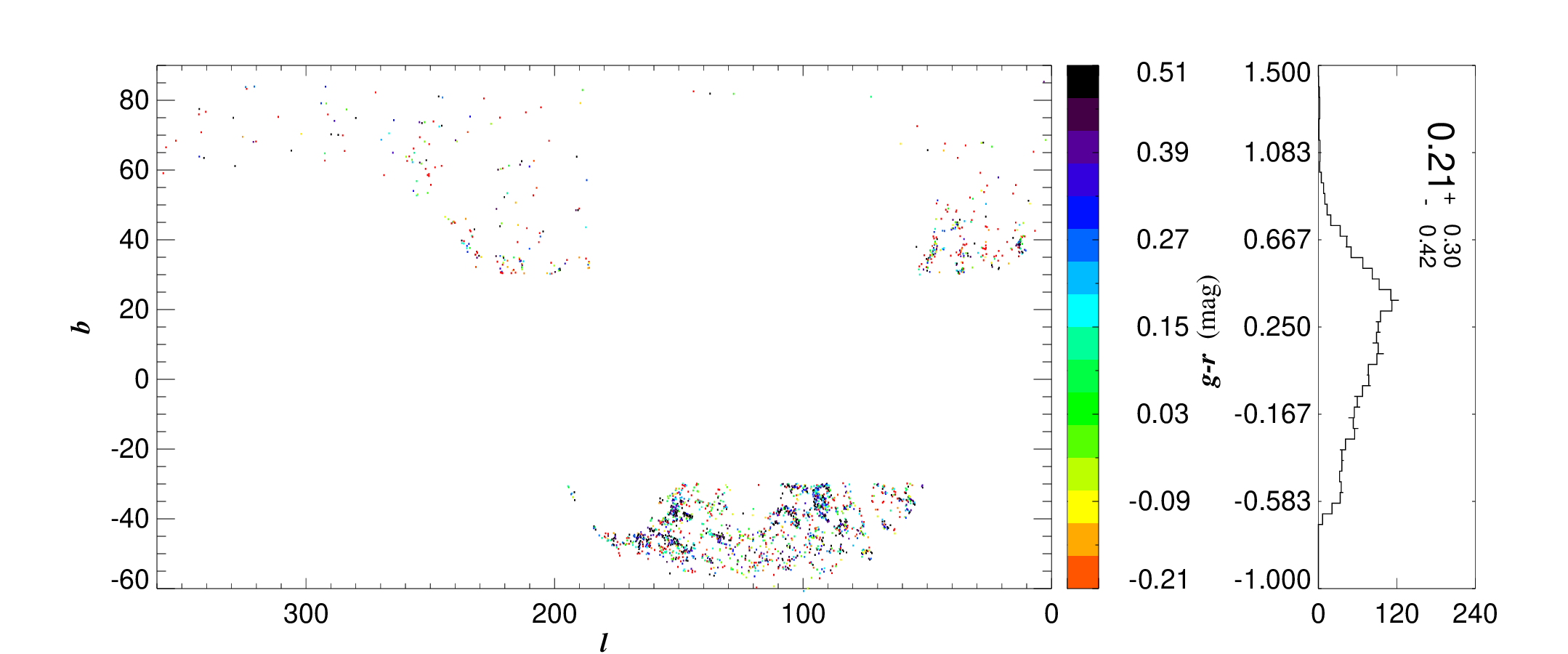}
        \label{fig:coverage_para_all_grid_g-r_color}        
        \end{figure*} 
  
        \begin{figure*}[htbp]
        \centering
        \includegraphics[width=0.9\linewidth]{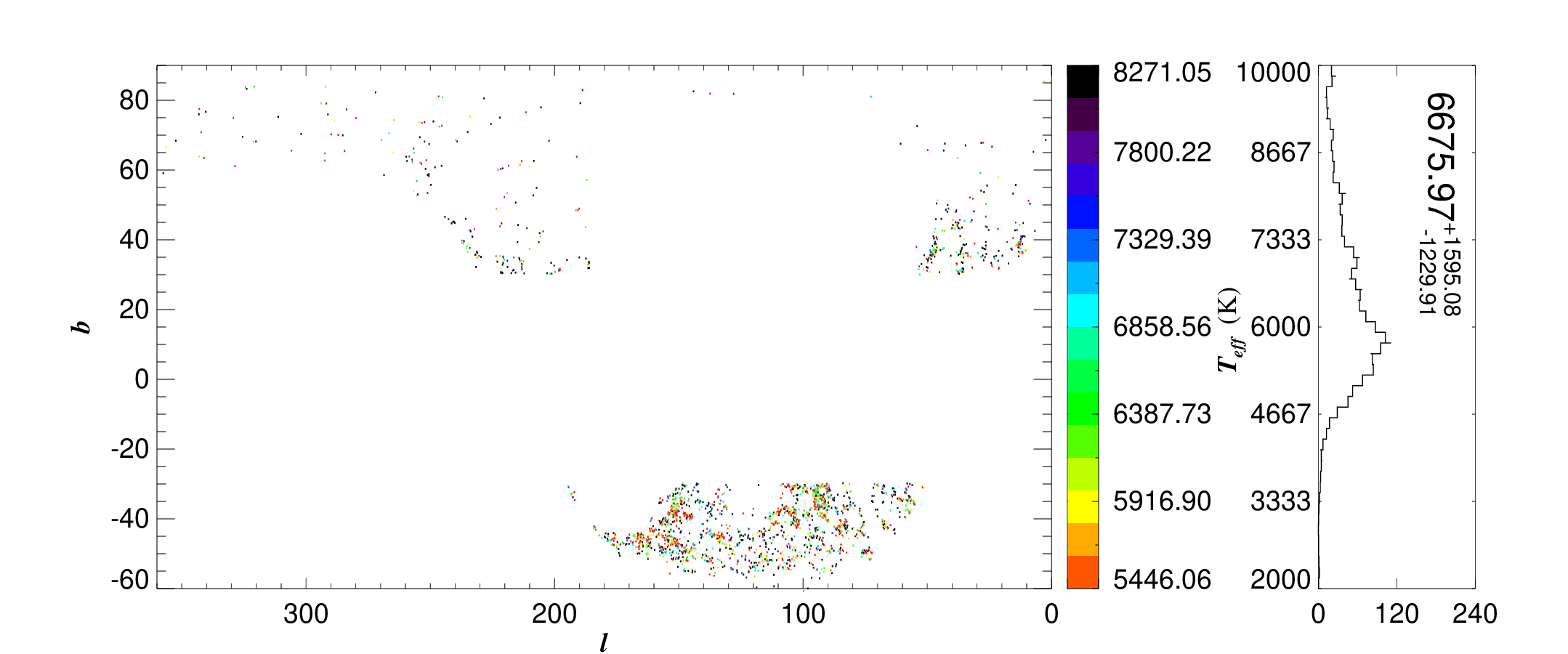}
        \label{fig:coverage_para_all_grid_Teff}        
        \end{figure*}   
        
        \begin{figure*}[htbp]
        \centering
        \includegraphics[width=0.9\linewidth]{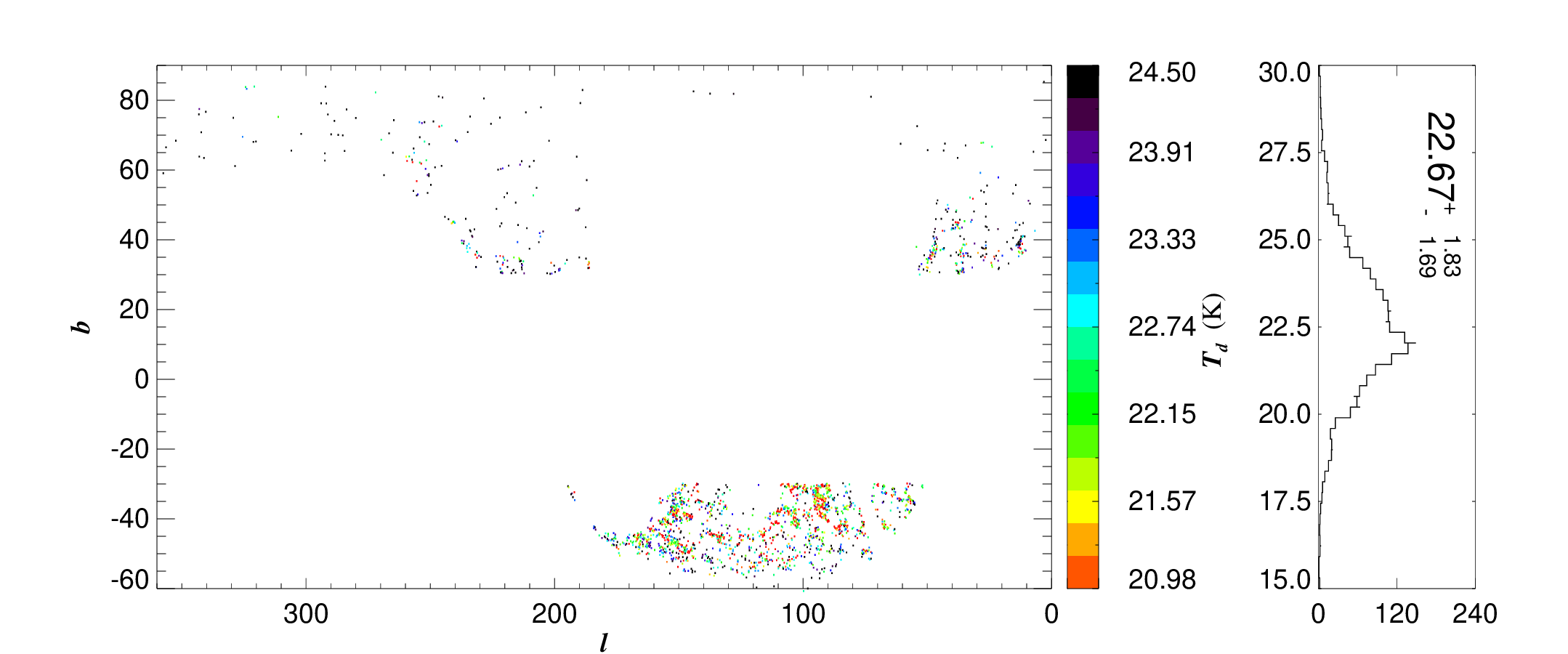}
        \label{fig:coverage_para_all_grid_Td}        
        \end{figure*}       

        \begin{figure*}[htbp]
        \centering
        \includegraphics[width=0.9\linewidth]{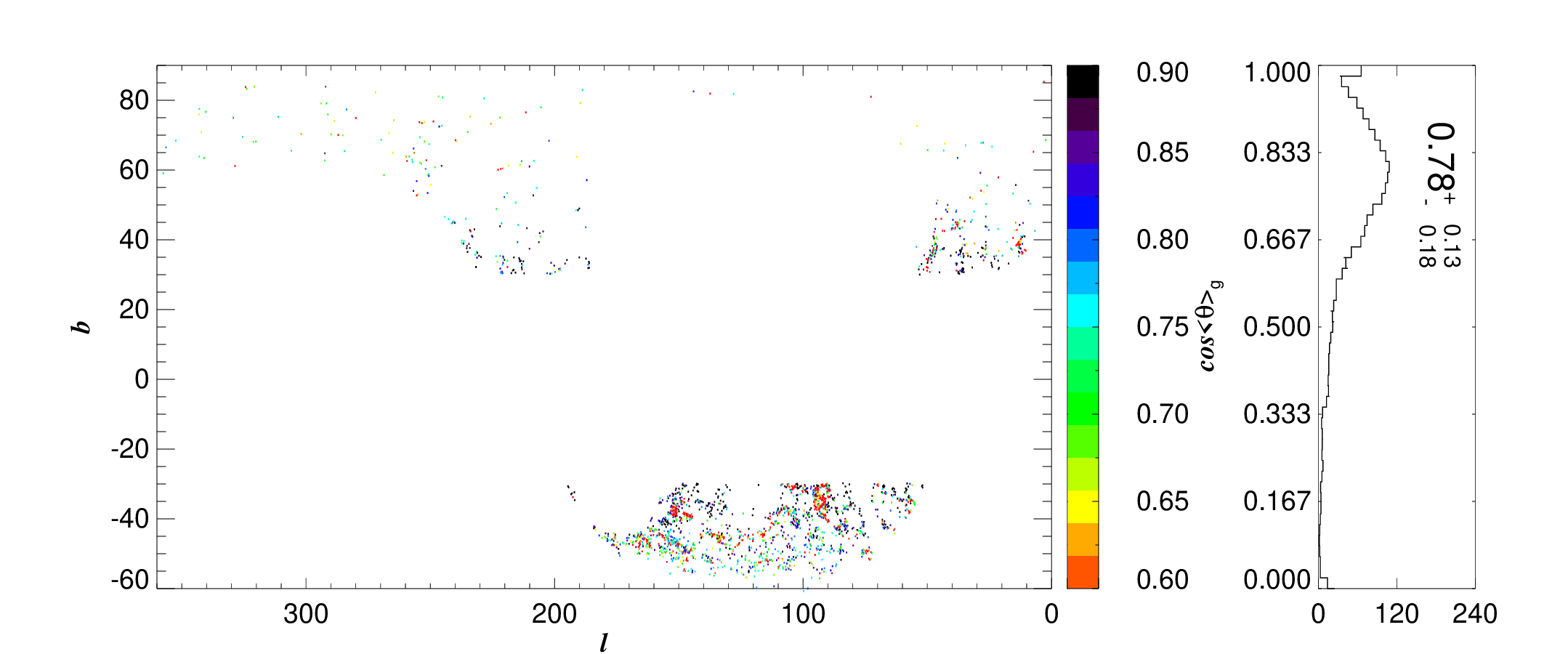}
        \label{fig:coverage_para_all_grid_asymmetry_g}        
        \end{figure*} 
        
        \begin{figure*}[htbp]
        \centering
        \includegraphics[width=0.9\linewidth]{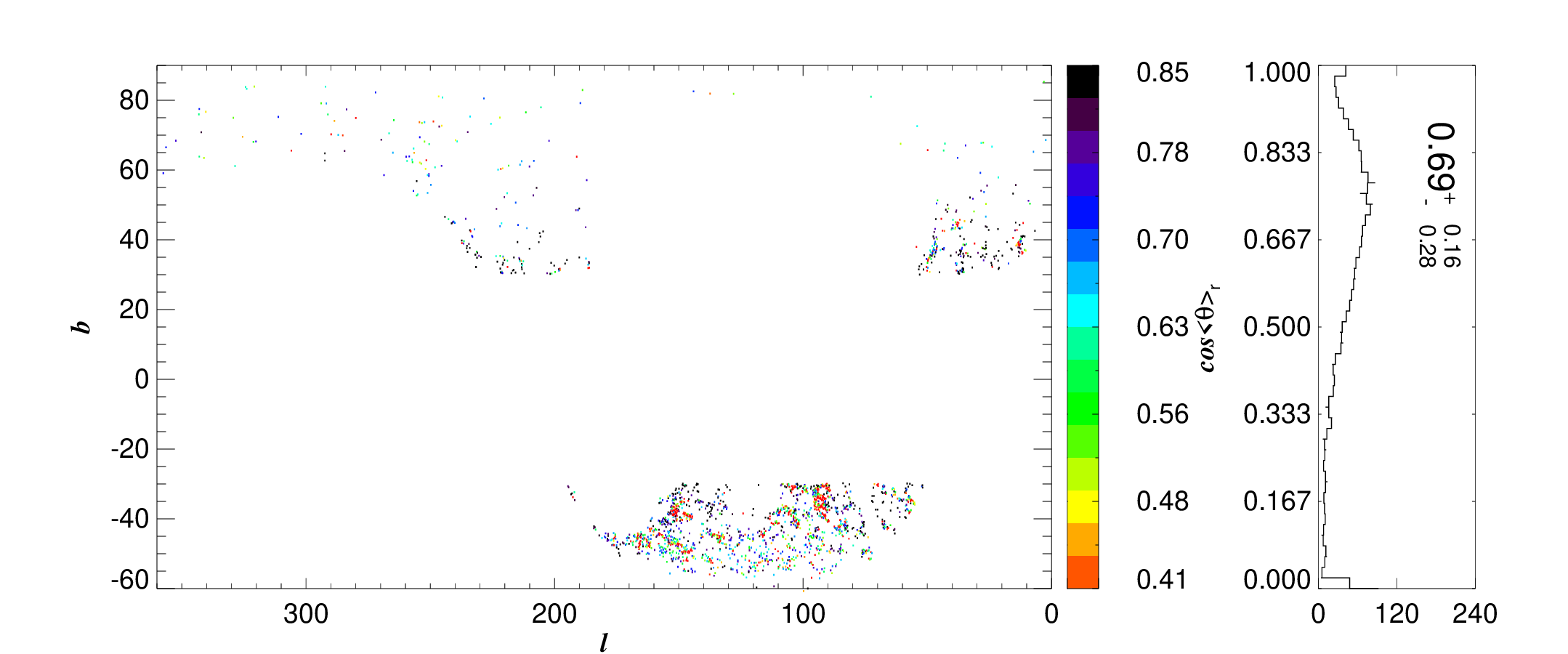}
        \label{fig:coverage_para_all_grid_asymmetry_r}        
        \end{figure*}

        \begin{figure*}[htbp]
        \centering
        \includegraphics[width=0.9\linewidth]{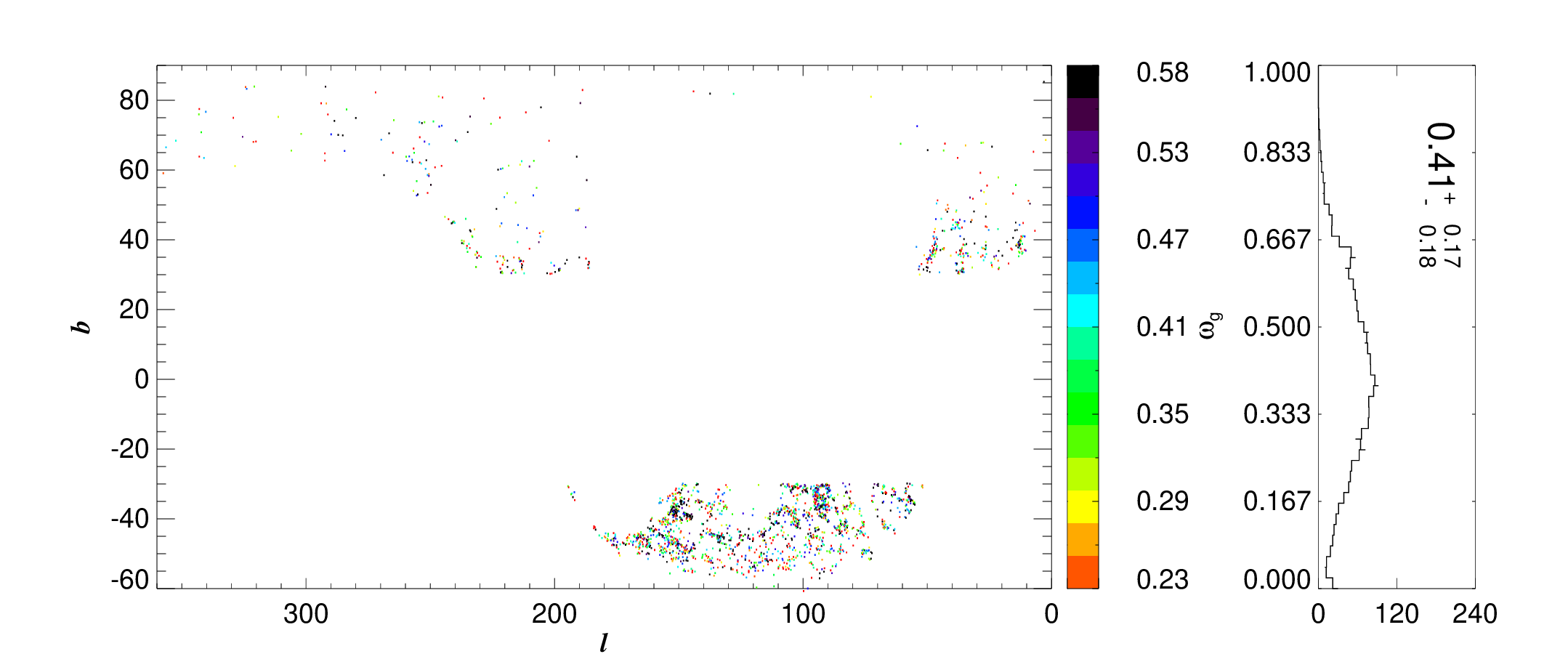}
        \label{fig:coverage_para_all_grid_albedo_g}        
        \end{figure*} 
        
\begin{sidewaystable}
\centering
\caption{Detailed information of identified cirri blocks. The full version of the table can be obtained from the online version of the article.} 
\label{table:cirri information}
\addtolength{\leftskip} {-1cm} 
\begin{tabular}{ccccccccccccccc} 
\hline
FID  & NUM & RA(J2000) & DEC(J2000) & GL    & GB          & $I_{100,med}$       & $I_{g,med}$          & $I_{r,med}$         & $S_{g,med}$           & $S_{r,med}$           & $k_g$                & $k_r$                 & $C_g$     & $C_r$ \\
     &     & (deg)     & (deg)      & (deg) & (deg)       &(MJy\,sr$^{-1}$)      &(kJy\,sr$^{-1}$)       & (kJy\, sr$^{-1}$) & (mag\, arcsec$^{-2}$)   & (mag\, arcsec$^{-2}$)   & ($\times$ 10$^{-3}$)      &($\times$ 10$^{-3}$)  &  &  \\ 
\hline
2607 & 1   & 19.75     & 17.16     & 132.30 & -45.21      & 5.36 $\pm$ 0.53    & 0.66 $\pm$ 0.65       & 0.82 $\pm$ 1.11     & 28.42 $\pm$ 0.92      & 28.19 $\pm$ 1.08      & 1.13 $\pm$ 0.13      & 1.86 $\pm$ 0.19       & 0.99      & 0.99  \\
2607 & 3   & 19.33     & 17.15     & 131.74 & -45.28      & 5.87 $\pm$ 0.93    & 1.05 $\pm$ 0.90       & 1.57 $\pm$ 1.51     & 27.92 $\pm$ 0.89      & 27.48 $\pm$ 1.22      & 0.88 $\pm$ 0.08      & 1.52 $\pm$ 0.17       & 0.99      & 0.96  \\
2607 & 4   & 19.34     & 16.75     & 131.83 & -45.67      & 6.10 $\pm$ 1.10    & 0.99 $\pm$ 0.87       & 1.56 $\pm$ 1.67     & 27.98 $\pm$ 0.88      & 27.49 $\pm$ 1.06      & 0.75 $\pm$ 0.09      & 1.32 $\pm$ 0.18       & 0.98      & 0.98  \\
2643 & 1   & 19.85     & 15.35     & 132.84 & -46.99      & 3.88 $\pm$ 0.40    & 0.29 $\pm$ 0.38       & 0.60 $\pm$ 0.65     & 29.30 $\pm$ 1.56      & 28.52 $\pm$ 1.19      & 0.26 $\pm$ 0.03      & 0.69 $\pm$ 0.21       & 0.83      & 0.87  \\
2643 & 3   & 19.44     & 15.35     & 132.27 & -47.05      & 3.72 $\pm$ 0.31    & 0.30 $\pm$ 0.30       & 0.54 $\pm$ 0.46     & 29.33 $\pm$ 1.15      & 28.66 $\pm$ 1.15      & 0.33 $\pm$ 0.12      & 0.79 $\pm$ 0.20       & 0.92      & 0.81  \\
2824 & 1   & 20.31     & 16.31     & 133.24 & -45.96      & 4.73 $\pm$ 0.36    & 0.61 $\pm$ 0.36       & 0.68 $\pm$ 0.60     & 28.52 $\pm$ 0.78      & 28.39 $\pm$ 1.58      & 0.32 $\pm$ 0.19      & 0.53 $\pm$ 0.17       & 0.87      & 0.93  \\
2824 & 2   & 20.31     & 15.92     & 133.34 & -46.35      & 4.52 $\pm$ 0.37    & 0.89 $\pm$ 0.58       & 1.06 $\pm$ 0.82     & 28.10 $\pm$ 0.91      & 27.91 $\pm$ 1.24      & 0.57 $\pm$ 0.33      & 0.29 $\pm$ 0.55       & 0.98      & 0.92  \\
3000 & 1   & 20.74     & 17.29     & 133.59 & -44.93      & 7.05 $\pm$ 0.85    & 1.39 $\pm$ 1.22       & 2.31 $\pm$ 2.64     & 27.62 $\pm$ 0.99      & 27.07 $\pm$ 1.07      & 0.78 $\pm$ 0.20      & 1.59 $\pm$ 0.33       & 0.98      & 0.98  \\
3000 & 2   & 20.75     & 16.89     & 133.69 & -45.32      & 6.87 $\pm$ 0.83    & 1.50 $\pm$ 1.16       & 2.55 $\pm$ 2.90     & 27.54 $\pm$ 0.89      & 26.96 $\pm$ 1.24      & 0.94 $\pm$ 0.19      & 1.94 $\pm$ 0.44       & 0.95      & 0.95  \\
3000 & 3   & 20.33     & 17.28     & 133.04 & -45.00      & 5.84 $\pm$ 0.35    & 0.61 $\pm$ 0.87       & 1.00 $\pm$ 1.45     & 28.51 $\pm$ 1.13      & 27.97 $\pm$ 1.18      & 1.19 $\pm$ 0.19      & 2.38 $\pm$ 0.35       & 0.67      & 0.73  \\
3000 & 4   & 20.33     & 16.89     & 133.13 & -45.39      & 5.43 $\pm$ 0.43    & 0.55 $\pm$ 0.34       & 0.66 $\pm$ 0.58     & 28.63 $\pm$ 0.99      & 28.42 $\pm$ 1.01      & 0.38 $\pm$ 0.15      & 0.59 $\pm$ 0.21       & 1.00      & 1.00  \\
3039 & 1   & 20.86     & 15.50     & 134.21 & -46.67      & 4.48 $\pm$ 0.34    & 0.64 $\pm$ 0.46       & 0.72 $\pm$ 0.76     & 28.49 $\pm$ 1.20      & 28.38 $\pm$ 1.39      & 1.14 $\pm$ 0.27      & 1.71 $\pm$ 0.46       & 0.98      & 1.00  \\
3039 & 3   & 20.45     & 15.50     & 133.63 & -46.74      & 5.30 $\pm$ 0.70    & 1.07 $\pm$ 0.78       & 1.88 $\pm$ 1.44     & 27.90 $\pm$ 0.88      & 27.29 $\pm$ 1.11      & 0.89 $\pm$ 0.14      & 1.57 $\pm$ 0.28       & 1.00      & 1.00  \\
3220 & 1   & 21.32     & 16.44     & 134.58 & -45.66      & 4.65 $\pm$ 0.47    & 0.23 $\pm$ 0.23       & 0.27 $\pm$ 0.32     & 29.67 $\pm$ 1.08      & 29.45 $\pm$ 1.00      & 0.26 $\pm$ 0.15      & 0.11 $\pm$ 0.18       & 0.81      & 0.68  \\
3220 & 3   & 20.90     & 16.44     & 134.02 & -45.74      & 5.19 $\pm$ 0.72    & 0.83 $\pm$ 0.63       & 1.34 $\pm$ 1.20     & 28.17 $\pm$ 1.06      & 27.65 $\pm$ 1.24      & 0.56 $\pm$ 0.15      & 1.09 $\pm$ 0.29       & 0.98      & 0.99  \\
3220 & 4   & 20.90     & 16.04     & 134.12 & -46.13      & 5.35 $\pm$ 0.64    & 0.52 $\pm$ 0.51       & 0.62 $\pm$ 0.83     & 28.68 $\pm$ 0.97      & 28.49 $\pm$ 1.55      & 0.38 $\pm$ 0.18      & 0.48 $\pm$ 0.27       & 0.98      & 0.95  \\
3387 & 1   & 21.76     & 17.39     & 134.92 & -44.65      & 6.50 $\pm$ 0.83    & 0.52 $\pm$ 0.43       & 0.75 $\pm$ 0.78     & 28.69 $\pm$ 0.95      & 28.28 $\pm$ 1.07      & 0.31 $\pm$ 0.07      & 0.65 $\pm$ 0.15       & 0.96      & 0.96  \\
3387 & 3   & 21.34     & 17.39     & 134.36 & -44.73      & 6.41 $\pm$ 0.61    & 1.07 $\pm$ 0.64       & 1.56 $\pm$ 1.16     & 27.90 $\pm$ 0.76      & 27.51 $\pm$ 1.08      & 0.70 $\pm$ 0.10      & 1.17 $\pm$ 0.14       & 1.00      & 1.00  \\
3387 & 4   & 21.34     & 16.99     & 134.47 & -45.12      & 6.04 $\pm$ 0.32    & 0.49 $\pm$ 0.35       & 0.80 $\pm$ 0.57     & 28.75 $\pm$ 1.03      & 28.21 $\pm$ 1.33      & 0.57 $\pm$ 0.11      & 0.97 $\pm$ 0.28       & 0.85      & 0.95  \\
3748 & 1   & 22.78     & 17.48     & 136.23 & -44.36      & 5.14 $\pm$ 0.31    & 0.78 $\pm$ 0.38       & 0.88 $\pm$ 0.78     & 28.25 $\pm$ 0.73      & 28.14 $\pm$ 1.25      & 0.69 $\pm$ 0.19      & 0.63 $\pm$ 0.31       & 0.99      & 0.85  \\
3748 & 2   & 22.78     & 17.08     & 136.34 & -44.75      & 5.51 $\pm$ 0.42    & 0.76 $\pm$ 0.81       & 0.94 $\pm$ 1.20     & 28.27 $\pm$ 1.15      & 28.04 $\pm$ 1.46      & 0.30 $\pm$ 0.34      & 0.48 $\pm$ 0.36       & 0.89      & 0.90  \\
3748 & 3   & 22.37     & 17.48     & 135.68 & -44.44      & 7.17 $\pm$ 1.41    & 1.22 $\pm$ 0.89       & 1.72 $\pm$ 1.58     & 27.75 $\pm$ 0.80      & 27.38 $\pm$ 1.15      & 0.45 $\pm$ 0.07      & 0.90 $\pm$ 0.10       & 0.99      & 0.99  \\
3748 & 4   & 22.36     & 17.09     & 135.79 & -44.84      & 5.90 $\pm$ 0.69    & 0.62 $\pm$ 0.47       & 0.72 $\pm$ 0.59     & 28.49 $\pm$ 0.80      & 28.34 $\pm$ 1.06      & 0.25 $\pm$ 0.19      & 0.42 $\pm$ 0.24       & 0.90      & 0.93  \\
3965 & 3   & 22.91     & 16.66     & 136.64 & -45.14      & 5.26 $\pm$ 0.34    & 0.27 $\pm$ 0.18       & 0.36 $\pm$ 0.35     & 29.41 $\pm$ 1.10      & 29.07 $\pm$ 1.30      & 0.02 $\pm$ 0.09      & 0.54 $\pm$ 0.18       & 0.84      & 0.99  \\
4100 & 2   & 23.77     & 17.17     & 137.61 & -44.45      & 4.90 $\pm$ 0.23    & 0.36 $\pm$ 0.20       & 0.44 $\pm$ 0.32     & 29.09 $\pm$ 0.94      & 28.86 $\pm$ 1.26      & 0.50 $\pm$ 0.13      & 0.65 $\pm$ 0.17       & 0.87      & 0.95  \\
4100 & 3   & 23.36     & 17.58     & 136.95 & -44.15      & 4.90 $\pm$ 0.28    & 0.73 $\pm$ 0.42       & 0.61 $\pm$ 0.63     & 28.33 $\pm$ 0.97      & 28.52 $\pm$ 1.29      & 0.64 $\pm$ 0.20      & 0.58 $\pm$ 0.43       & 0.87      & 0.87  \\
4100 & 4   & 23.36     & 17.18     & 137.07 & -44.54      & 4.99 $\pm$ 0.34    & 0.52 $\pm$ 0.36       & 0.71 $\pm$ 0.53     & 28.68 $\pm$ 0.92      & 28.35 $\pm$ 0.97      & 0.41 $\pm$ 0.16      & 1.09 $\pm$ 0.25       & 0.94      & 0.96  \\
4490 & 2   & 24.82     & 17.23     & 138.94 & -44.15      & 6.18 $\pm$ 1.09    & 0.50 $\pm$ 0.50       & 0.64 $\pm$ 0.86     & 28.72 $\pm$ 0.92      & 28.45 $\pm$ 1.48      & 0.38 $\pm$ 0.07      & 0.69 $\pm$ 0.08       & 0.93      & 0.99  \\
4490 & 3   & 24.41     & 17.64     & 138.28 & -43.85      & 5.06 $\pm$ 0.29    & 0.33 $\pm$ 0.21       & 0.41 $\pm$ 0.34     & 29.18 $\pm$ 0.77      & 28.99 $\pm$ 1.20      & 0.22 $\pm$ 0.12      & 0.27 $\pm$ 0.16       & 0.89      & 0.76  \\
\hline
\end{tabular}
\end{sidewaystable}

\begin{table}[]
\centering
\addtolength{\leftskip} {-2cm} 
\caption{Detailed parameters of identified cirri blocks. The full version of the table can be obtained from the online version of the article.}
\label{table:cirri parameter}
\begin{tabular}{ccccccccc}

\hline
FID  & NUM   & $g-r$                  & $T_d$                    & $T_{eff}$                  & cos$<\theta>_g$ & cos$<\theta>_r$ & $\omega_g$ & $\omega_r$ \\
     &       & (mag)                  & (K)                      & (K)                  & &  &  &  \\
\hline
2607 & 1     & 0.56  $\pm$ 0.02       & 21.49 $\pm$ 0.36         & 5377.06  $\pm$ 59.72       & 0.61            & 0.48            & 0.71     & 0.79      \\
2607 & 3     & 0.53  $\pm$ 0.02       & 20.31 $\pm$ 0.19         & 5477.42  $\pm$ 79.50       & 0.32            & 0.00            & 0.67     & 0.76      \\
2607 & 4     & 0.68  $\pm$ 0.02       & 20.38 $\pm$ 0.18         & 5037.05  $\pm$ 51.71       & 0.36            & 0.00            & 0.63     & 0.74      \\
2643 & 1     & 0.39  $\pm$ 0.07       & 21.32 $\pm$ 0.81         & 5973.65  $\pm$ 277.48      & 0.86            & 0.64            & 0.34     & 0.58      \\
2643 & 3     & 0.16  $\pm$ 0.10       & 23.00 $\pm$ 0.81         & 7018.55  $\pm$ 553.84      & 0.85            & 0.68            & 0.40     & 0.61      \\
2824 & 1     & 0.38  $\pm$ 0.06       & 21.34 $\pm$ 0.67         & 5990.84  $\pm$ 235.38      & 0.64            & 0.59            & 0.40     & 0.52      \\
2824 & 2     & 0.31  $\pm$ 0.05       & 22.17 $\pm$ 0.71         & 6265.16  $\pm$ 223.66      & 0.43            & 0.31            & 0.55     & 0.36      \\
3000 & 1     & 0.74  $\pm$ 0.04       & 20.33 $\pm$ 0.32         & 4884.77  $\pm$ 87.92       & 0.08            & 0.00            & 0.65     & 0.78      \\
3000 & 2     & 0.96  $\pm$ 0.02       & 19.25 $\pm$ 0.35         & 4385.54  $\pm$ 34.71       & 0.00            & 0.00            & 0.69     & 0.81      \\
3000 & 3     & 0.53  $\pm$ 0.02       & 20.35 $\pm$ 0.77         & 5483.64  $\pm$ 60.52       & 0.65            & 0.36            & 0.73     & 0.83      \\
3000 & 4     & 0.41  $\pm$ 0.06       & 22.42 $\pm$ 0.47         & 5889.56  $\pm$ 213.81      & 0.69            & 0.61            & 0.45     & 0.55      \\
3039 & 1     & 0.36  $\pm$ 0.11       & 22.35 $\pm$ 0.86         & 6057.03  $\pm$ 417.44      & 0.62            & 0.56            & 0.71     & 0.77      \\
3039 & 3     & 0.63  $\pm$ 0.03       & 20.34 $\pm$ 0.37         & 5180.12  $\pm$ 79.48       & 0.30            & 0.00            & 0.66     & 0.76      \\
3220 & 1     & -0.13 $\pm$ 0.27       & 22.95 $\pm$ 0.92         & 9147.81  $\pm$ 2600.53     & 0.91            & 0.88            & 0.35     & 0.18      \\
3220 & 3     & 0.64  $\pm$ 0.03       & 22.65 $\pm$ 0.24         & 5148.10  $\pm$ 89.32       & 0.47            & 0.11            & 0.55     & 0.69      \\
3220 & 4     & 0.39  $\pm$ 0.06       & 20.73 $\pm$ 0.42         & 5970.02  $\pm$ 214.52      & 0.70            & 0.63            & 0.45     & 0.49      \\
3387 & 1     & 0.44  $\pm$ 0.06       & 19.87 $\pm$ 0.32         & 5774.36  $\pm$ 222.49      & 0.71            & 0.54            & 0.42     & 0.58      \\
3387 & 3     & 0.59  $\pm$ 0.03       & 21.95 $\pm$ 0.24         & 5298.83  $\pm$ 82.18       & 0.30            & 0.00            & 0.62     & 0.72      \\
3387 & 4     & 0.41  $\pm$ 0.05       & 19.12 $\pm$ 0.87         & 5875.13  $\pm$ 199.83      & 0.73            & 0.51            & 0.57     & 0.67      \\
3748 & 1     & 0.43  $\pm$ 0.11       & 19.96 $\pm$ 1.09         & 5810.56  $\pm$ 388.95      & 0.53            & 0.45            & 0.60     & 0.56      \\
3748 & 2     & 0.22  $\pm$ 0.07       & 24.73 $\pm$ 0.68         & 6685.05  $\pm$ 333.37      & 0.54            & 0.40            & 0.40     & 0.49      \\
3748 & 3     & 0.54  $\pm$ 0.04       & 20.18 $\pm$ 0.18         & 5451.83  $\pm$ 110.70      & 0.20            & 0.00            & 0.53     & 0.67      \\
3748 & 4     & 0.10  $\pm$ 0.06       & 22.74 $\pm$ 0.25         & 7352.50  $\pm$ 335.15      & 0.64            & 0.56            & 0.36     & 0.46      \\
3965 & 3     & -0.42 $\pm$ 0.33       & 21.55 $\pm$ 0.81         & 13306.45 $\pm$ 7361.66     & 0.89            & 0.82            & 0.04     & 0.53      \\
4100 & 2     & 0.20  $\pm$ 0.11       & 15.14 $\pm$ 4.77         & 6799.49  $\pm$ 534.45      & 0.83            & 0.77            & 0.52     & 0.57      \\
4100 & 3     & 0.15  $\pm$ 0.09       & 15.43 $\pm$ 2.36         & 7067.38  $\pm$ 505.79      & 0.57            & 0.65            & 0.58     & 0.54      \\
4100 & 4     & 0.16  $\pm$ 0.08       & 15.25 $\pm$ 1.17         & 7033.07  $\pm$ 413.07      & 0.71            & 0.58            & 0.46     & 0.69      \\
4490 & 2     & 0.54  $\pm$ 0.03       & 23.48 $\pm$ 0.26         & 5435.29  $\pm$ 97.75       & 0.73            & 0.63            & 0.46     & 0.60      \\
4490 & 3     & 0.12  $\pm$ 0.10       & 16.21 $\pm$ 1.74         & 7227.38  $\pm$ 564.04      & 0.85            & 0.79            & 0.32     & 0.35      \\
\hline
\end{tabular}
\end{table}

\clearpage
\bibliographystyle{aasjournal}
\bibliography{ref}

\end{document}